\documentstyle[seceq,preprint]{jpsj}

\newcommand{\ve}{\mbox{$\varepsilon$}}

\def\runtitle{Numerical Renormalization Group Study of non-Fermi-liquid State
on Dilute Uranium Systems}
\def\runauthor{Yukihiro {\sc Shimizu}, Osamu {\sc Sakai} and Shunya {\sc Suzuki}}

\title
{
Numerical Renormalization Group Study of non-Fermi-liquid State
on Dilute Uranium Systems
}

\author
{ 
Yukihiro {\sc Shimizu}\footnote{E-mail: yshmz@snow.apph.tohoku.ac.jp},
Osamu {\sc Sakai}$^{1}$
and Shunya {\sc Suzuki}$^{1}$
}

\inst
{
Department of Applied Physics, Tohoku University, Sendai 980-8579\\
$^1$Department of Physics, Tohoku University, Sendai 980-8578
}

\recdate
{
\today
}

\abst
{
We investigate the non-Fermi-liquid (NFL) behavior
of the impurity Anderson model (IAM)
with non-Kramers doublet ground state of the f$^2$ configuration
under the tetragonal crystalline electric field (CEF).
The low energy spectrum 
is explained by a combination of the NFL
and the local-Fermi-liquid parts which are independent with each other.
The NFL part of the spectrum has the same form to that of
two-channel-Kondo model (TCKM).
We have a parameter range that the IAM shows the $- \ln T$ divergence
of the magnetic susceptibility together with the positive magneto resistance.
We point out a possibility that the anomalous properties of
U$_x$Th$_{1-x}$Ru$_2$Si$_2$ including the decreasing resistivity
with decreasing temperature can be explained by the NFL scenario of the TCKM type.
We also investigate an effect of the lowering of the crystal symmetry.
It breaks the NFL behavior at around the temperature, $\delta /10$,
where $\delta$ is the orthorhombic CEF splitting.
The NFL behavior is still
expected above the temperature, $\delta/10$.
}

\kword
{
non-Fermi-liquid, impurity Anderson model, dilute Uranium compound,
magnetization, magneto resistance, numerical renormalization group
}

\begin{document}
\sloppy
\maketitle

\input epsf

\section{Introduction}
The material, U$_x$Th$_{1-x}$Ru$_2$Si$_2$, seems to show the 
non-Fermi-liquid (NFL) behavior even in
the dilute uranium limit.~\cite{rf:ami1,rf:ami2,rf:ami3}
Its magnetic susceptibility and $\gamma$-coefficient
of the specific heat show the $-\ln T$ divergence at the low temperatures.
At the same time the electrical resistivity decreases
with decreasing temperature.
Particular interests have been aroused
in the NFL behavior driven by the single site effect.~\cite{rf:cox1}
This material has the tetragonal type structure,
and the valence of the uranium ion is expected to be mainly, U$^{4+}$ (5f$^2$).
The crystalline electric field (CEF) ground state is also expected to be
the non-Kramers doublet state from the measurement of the magnetization.

Many theoretical studies have been done for the NFL behavior
of the dilute uranium compounds.
Cox has pointed out the possibility that
the electronic state of ion with f$^2$ configuration in a specific situation
can be mapped on the $S$=1/2-two-channel Kondo model
($S$=1/2-TCKM),~\cite{rf:cox2, rf:cox3}
which is known to show the NFL
behavior.~\cite{rf:nozieres, rf:bethe1, rf:bethe2, rf:cft}
The required situation is that
the lowest CEF state of the f$^2$ configuration is the non-Kramers doublet state,
and two kinds of conduction electrons, both of which have doubly degenerate time
reversal pair, can hybridize with the f state.
However, there are two remained problems in the application of the $S$=1/2-TCKM
to the NFL behavior of U$_x$Th$_{1-x}$Ru$_2$Si$_2$:
First, the $S$=1/2-TCKM can not explain the decrease of the resistivity
with decreasing temperature.
Usually opposite temperature dependence is expected when the coefficient
of the $-\ln T$ term of the magnetic susceptibility
is sizable.~\cite{rf:ours1, rf:kusunose}
Secondly, the mapping is done assuming very restricted situation.
It is not clear whether the NFL state is stable or not
when the realistic situation of the  CEF states is taken into account.
For the first problem we have modified the TCKM,
and have proposed the extended two-channel Anderson model
(ETCAM) that shows similar behaviors of the resistivity
and the magnetic susceptibility
to those of the experiment.~\cite{rf:ours2, rf:ours3}
However, applicability of the ETCAM to the uranium problem is not obvious
at present, because the intimate one-to-one mapping of electronic state has not
been verified.
In this paper, we check whether the anomalous properties
of U$_x$Th$_{1-x}$Ru$_2$Si$_2$ can be explained
or not based on a more realistic model.

For the second problem Koga and Shiba have studied the stability of the NFL state
by taking into account the first excited state
of the f$^2$ configuration.~\cite{rf:koga1, rf:koga2, rf:koga3}
They applied the numerical renormalization group (NRG) method,
and  has shown that the NFL state is still stable, if the CEF splitting is large.
In previous paper, we have also studied the stability of the NFL state
based directly on the impurity Anderson model (IAM),~\cite{rf:ours4}
because preceding studies have been done applying the Schrieffer-Wolff
transformation by assuming the strong correlation limit.
It has been shown that the ground state property can change to the NFL state
from the usual Kondo type singlet state when the intensity of the hybridization
is gradually weakened,
even though all CEF states are taken into account,
and the valence is apart from f$^2$.
But details of the low energy properties of the NFL state have not been clarified.

The first purpose of this paper is to analyze the low energy spectrum
of the IAM given by the NRG method in detail.~\cite{rf:ours5}
We examine the model that the single electron orbitals split into
three doublets under the tetragonal CEF.
The CEF ground state is the non-Kramers doublet of f$^2$ configuration.
In \S 3 we show that the low energy spectrum is explained
by a combination of two components which are given by
the NFL and the local-Fermi-liquid (LFL) fixed point Hamiltonians,
and which are independent with each other.
The NFL part of the low energy spectrum has the same form to that of the TCKM.
The second purpose is to consider whether the anomalous behavior
in U$_x$Th$_{1-x}$Ru$_2$Si$_2$ is explained from the IAM or not.
In \S 4 and 5 we show that the large $-\ln T$ term of the magnetic
susceptibility can be expected together with the positive magneto resistance
at very low temperature, when the hybridization is not so weak.
The positive magneto resistance indicates that the temperature dependence of
the resistivity decreases with decreasing temperature.
The third purpose is to study the effect of the releasing process of the
residual entropy which is inherent of the $S$=1/2-TCKM.
In \S 6 we examine the NFL behavior in the magnetization
by breaking the NFL state with lowering the crystal symmetry.

\section{Model}
We consider the following IAM under tetragonal CEF,
\begin{eqnarray}
\label{eq:hamiltonian}
{\cal H} &=& {\cal H}_f + {\cal H}_{cf} + {\cal H}_c, \\
{\cal H}_f &=& \sum_{\Gamma \gamma}
\ve(\Gamma) n_{f \Gamma \gamma}
+ {U \over 2} \sum_{(\Gamma \gamma) \ne (\Gamma' \gamma')}
n_{f \Gamma \gamma}n_{f \Gamma' \gamma'} \nonumber \\
&-& {I \over 49} \sum_{\{\Gamma \gamma\}}
\left[{\mib j}_{\Gamma \gamma \Gamma' \gamma'}
{\mib j}_{\Gamma'' \gamma'' \Gamma''' \gamma'''}
f^\dagger_{\Gamma \gamma}f_{\Gamma' \gamma'}
f^\dagger_{\Gamma'' \gamma''}f_{\Gamma''' \gamma'''} \right. \nonumber \\
& & \left. -{35 n_f \over 4}
\right], \\
{\cal H}_{cf} &=& \sum_k \sum_{\Gamma \gamma} \left(
Vf^\dagger_{\Gamma \gamma}c_{k \Gamma \gamma} + {\rm h.c.} \right), \\
{\cal H}_c &=& \sum_k \sum_{\Gamma \gamma}
\ve_k c^\dagger_{k \Gamma \gamma}c_{k \Gamma \gamma},
\end{eqnarray}
where $f_{\Gamma \gamma}$ ($c_{k \Gamma \gamma}$)
is the annihilation operator of f-electron
with the $\gamma$-th component of $\Gamma$-irreducible representation
(conduction electron with wave number $k$),
and $\ve(\Gamma)$, $U$ and $I$ denote the single f-electron energy,
Coulomb and exchange interaction constants, respectively.
For simplicity we assume the large spin-orbit interaction of f-electron,
so only the $j=5/2$ orbitals are considered.
The orbitals split into a quartet, $\Gamma_8$, and a doublet,
$|f^1 \Gamma_7^{(2)}, \pm \rangle$ = $\pm \sqrt{1/6} | \pm 5/2 \rangle$
$\mp \sqrt{5/6} | \mp 3/2 \rangle$ under the cubic CEF
where $m$ in $| m \rangle$ of the right hand side of the equation
is the magnetic quantum number of $j$.~\cite{rf:butler}
The quartet splits into two doublets again,
$|f^1 \Gamma_7^{(1)}, \pm \rangle$ = $\pm \sqrt{5/6} | \pm 5/2 \rangle$
$\pm \sqrt{1/6} | \mp 3/2 \rangle$ and
$|f^1 \Gamma_6, \pm \rangle$ = $| \pm 1/2 \rangle$
under the tetragonal CEF.
The Coulomb and the exchange interactions are assumed to be the $j$-$j$ coupling
type.
Usually, the multiplet structure of U ion is approximated by the $L$-$S$
coupling scheme.
So the present $j$-$j$ coupling scheme seems to be not applicable to U ion
at first glance.
However, we note the ground multiplets have same total angular momentum
for both coupling schemes.
We expect qualitative features of the low energy states within the ground
multiplet will be not so changed.
It is assumed
the f-electron hybridizes with the conduction electron which has the
same component of the irreducible representation.
The quantities, $V$ and $\ve_k$ denote the hybridization matrix and
the band energy, respectively.
It is also assumed the band is extending in energy from $-D$ to $D$ with
constant hybridization matrix, $\Gamma = \pi V^2/2D$.
The energy unit and the origin of the energy are chosen as $D=1$ and
the Fermi level, respectively.

We rewrite the Hamiltonian into a form to suite the NRG calculation.
First the conduction band is discretized by the logarithmic mesh to give good
sampling to states near the Fermi energy.~\cite{rf:nrg}
Next the Hamiltonian is transformed into an expression given by the shell orbits,
\begin{eqnarray}
{\cal H} &=& \lim_{L \rightarrow \infty} {\cal H}_L, \\
{\cal H}_L &=& {\cal H}_f + \sum_{\Gamma \gamma}
\left( \sqrt{A_\Lambda}V f^\dagger_{\Gamma \gamma}
s_{0 \Gamma \gamma} + {\rm h.c.} \right) + {\cal H}_L^0, \\
{\cal H}_L^0 &=& \sum_{\ell=0}^{L-1} \sum_{\Gamma \gamma} t_\ell
\left(  s^\dagger_{\ell+1 \Gamma \gamma}
s_{\ell \Gamma \gamma} + {\rm h.c.} \right),
\end{eqnarray}
where $s_{\ell \Gamma \gamma}$ is the annihilation operator of $\ell$-th shell
state with the $\gamma$-th component of the $\Gamma$-irreducible representation.
The hopping energy, $t_\ell$, between shell states is given by
$t_\ell = D(1+ \Lambda^{-1})\Lambda^{-\ell/2} \xi_\ell /2$,
where $\Lambda(>1)$ is the discretization parameter,
and $\xi_\ell$ tend to 1 when $\ell$ increases.
The quantity $A_\Lambda$ is the correction factor of order 1
for the discretization.
We first diagonalize the ${\cal H}_f$ term, and then by adding
the shell state successively from $\ell=0$, we diagonalize the
series of cluster Hamiltonian, $\{{\cal H}_L\}$, recursively.
At each step we retain about the 500 lower energy states to the next step.
This number is not so large, but the obtained eigen states in the low energy
region seem to have enough accuracy for qualitative discussions.
When the only 300 states are retained at each step,
the low energy eigen states are not changed essentially.

The valence of U ion in U$_x$Th$_{1-x}$Ru$_2$Si$_2$ is not so clear,
so we consider two cases: 
one is the case that
the valence fluctuation between 5f$^2$ and 5f$^1$ configurations
becomes dominant,
and the other is that between 5f$^2$ and 5f$^3$ configurations.
We call the former a f$^2$ - f$^1$ dominant fluctuation case,
and the latter a f$^2$ - f$^3$ dominant fluctuation case hereafter.

\section{Analysis of the Low Energy Fixed Point}
\subsection{Flow chart of energy levels of f$^2$ - f$^1$ dominant fluctuation case}
We first consider the f$^2$ - f$^1$ dominant fluctuation case.
We choose the parameters;
$\ve(\Gamma_7^{(1)})=-0.9$,
$\ve(\Gamma_6)=-0.75$, $\ve(\Gamma_7^{(2)})=-0.5$, $U=0.6$ and $I=8$.
The lowest CEF state becomes the non-Kramers doublet of the f$^2$ configuration
which has mainly the character of
$J=4$, $\Gamma_5$-irreducible representation of D$_4$-group,
and the energy of the state is $-1.446$.
The state is given as
$|f^2 \Gamma_5, \pm \rangle$ = $a | \pm 1/2, \mp 3/2 \rangle$
+ $b | \pm 3/2, \mp 5/2 \rangle$ + $c | \pm 5/2, \pm 1/2 \rangle$,
where $a=-0.3395$, $b=-0.1874$ and $c=0.9218$.
We have used the notation,
$ | m, m' \rangle \equiv f^\dagger_m f^\dagger_{m'} | 0 \rangle$,
where $m$ and $m'$ are the magnetic quantum numbers of $j$.
The excited states from the first to the forth are the singlet
of the f$^2$ configuration with the energies, -1.442, -1.286, -1.208 and -1.167,
respectively.
These states have mainly the character of $J=4$, and
are the $\Gamma_4$, $\Gamma_1$, $\Gamma_2$
and $\Gamma_3$-irreducible representations of D$_4$-group, respectively.
The other configurations, f$^0$ and from f$^3$ to f$^6$ have higher energy.

In our previous work we have shown that three types of the ground state;
the doublet, the CEF-singlet-like and the f$^0$-singlet-like ground states
appear successively when the hybridization strength is increased
by fixing the energy of the CEF states.~\cite{rf:ours4}
The low energy spectrum of the last two follows the LFL
theory, however, the first one can not be explained by the LFL theory.
In this work we concentrate on the weak hybridization case
and analyze the spectrum of the NFL state.
In Fig.~\ref{fig:1} the lower eigen energies
of each cluster Hamiltonian, ${\cal H}_L$, for the odd NRG step, $L$, are shown.
The energies are renormalized by $t_{L-1}$ at each step.
The renormalized energy levels of the states which has the same charge, $Q$,
and the magnetic quantum number, $M$, change smoothly,
and we call the figure
the flow chart of the renormalized energy levels (FCEL) hereafter.
The quantity, $M$, is defined by using modulo 4 as defined in the caption
of Fig.~\ref{fig:1} because of the tetragonal symmetry.
The renormalized energy levels tend to fixed values as $L$ increases
beyond a step $L \sim 23$.
The right hand side of the figure gives the energy spectrum
at the low energy fixed point,
and the hopping energy, $t_{L-1}$, at $L \sim 23$ indicates the measure
of the energy scale for the cross over that the system goes into the low energy
fixed point.
Each state is indicated by $i$ (the sequential number of states
from low to high energy), $Q$ and $M$.
The ground state ($i=1, 2$) is doublet,
and the first excited states are two doublet states
with one particle excitation ($i=3, 5$)
and one hole excitation ($i=4, 6$).
The second excited states ($i=7 \sim 10$) are the one hole excitation,
and they have larger energy than twice energy of the first excited states.
Therefore these low energy states can not be explained by the LFL theory.
\begin{fullfigure}
\epsfxsize=.9\linewidth
\epsfbox{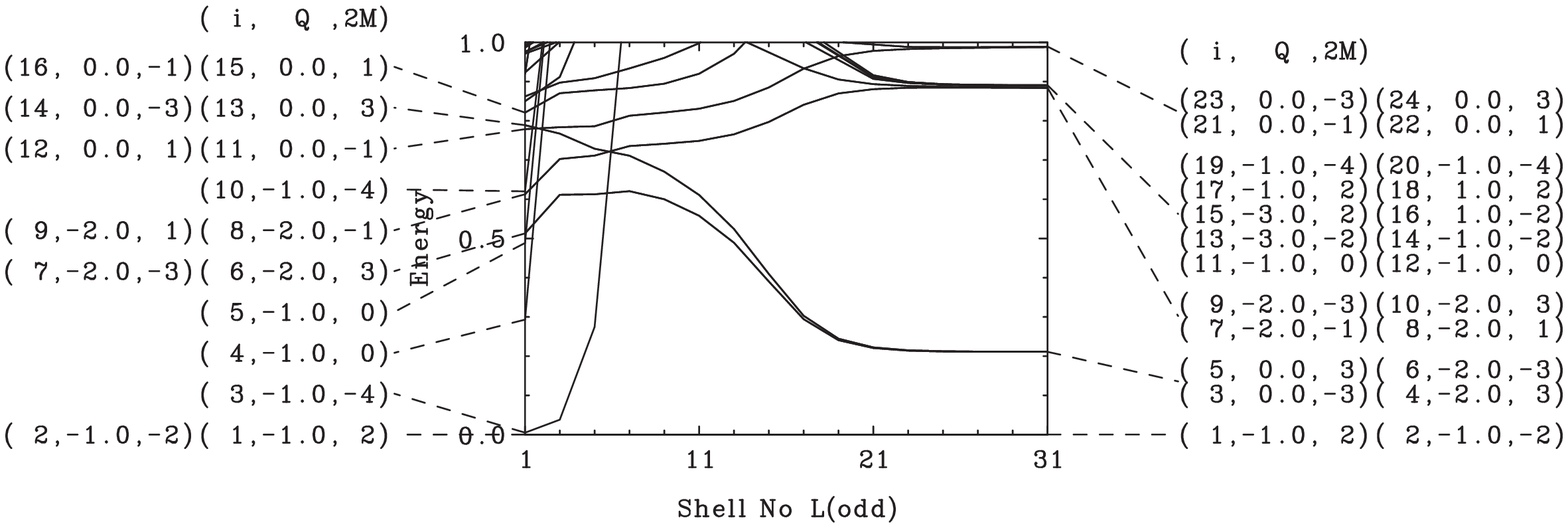}
\caption{
FCEL of the f$^2$ - f$^1$ dominant fluctuation case
for the odd renormalization step, $L$.
(In this case the number of shell orbits for the conduction electrons
is even, $L+1$. For example, number of shell orbits is two for $L=1$.)
The energies are renormalized by $t_{L-1}$ at each step.
The discretization parameter $\Lambda=8$ is used,
and about 500 states are retained at each step.
The parameters of IAM are;
$\ve(\Gamma_7^{(1)})=-0.9$,
$\ve(\Gamma_6)=-0.75$, $\ve(\Gamma_7^{(2)})=-0.5$,
$U=0.6$, $I=8$ and $V^2/2=0.035$.
The occupation number of the f-electron is 1.92.
Each state is labeled by the index, $i$ (sequential number of states from
low to high energy).
The charge of each state is denoted by $Q$,
which is defined as total electron number minus half of total orbital number,
$3(L+2)$.
The magnetic quantum number, $M$ is the index for grouping
of total magnetic quantum number, $J_z$.
When the quantity, $2J_z$ changes from $-4 + 8n$ to
$3 + 8n$, where $n$ is some integer, the quantity $2M$ varies from $-4$ to $3$.
The renormalized energies of states which has the same $Q$ and $M$
are connected as $L$ changed.
}
\label{fig:1}
\end{fullfigure}

\begin{figure}
\epsfxsize=.4\linewidth
\epsfbox{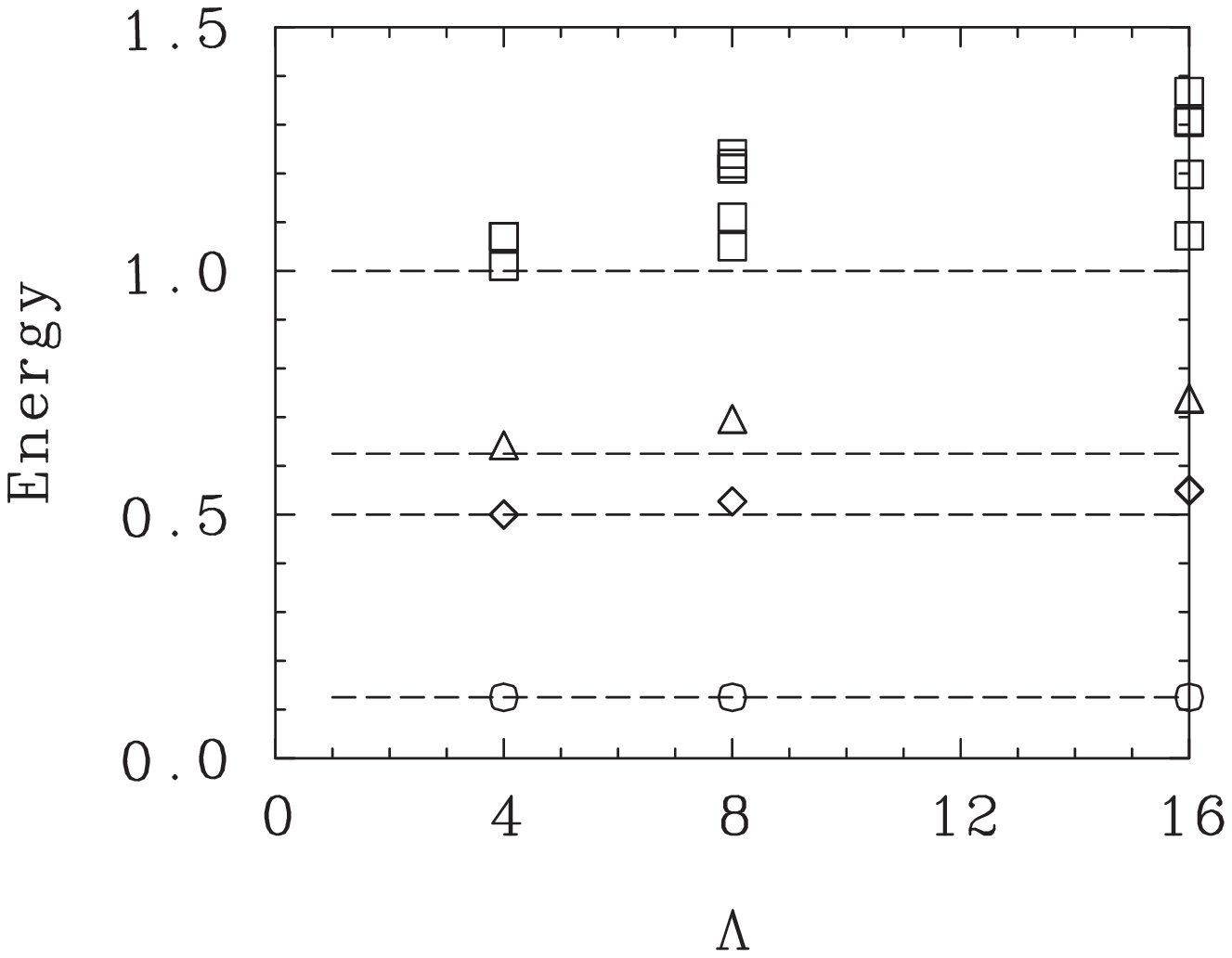}
\caption{
$\Lambda$ dependence of the energy
belongs to the NFL part at the low energy fixed point.
The symbols denote the renormalized energy, $r^* E$, given by the NRG calculation.
The parameters are the same as those in Fig.~\ref{fig:1} except $V$.
The hybridization width is chosen as $V^2/2=0.0433$, 0.0350 and 0.0310
for the cases of $\Lambda=4$, 8 and 16, respectively.
The dashed lines shows the energies which are expected by the CFT
for the $S$=1/2-TCKM.
The factor, $r^*$, is chosen so each energy of the first excited state
agrees with 1/8; $r^*=0.5787$, 0.5924 and 0.619
for $\Lambda=4$, 8 and 16, respectively.
}
\label{fig:2}
\end{figure}

We compare the low energy states with those of $S$=1/2-TCKM which are
derived by the conformal field theory (CFT).~\cite{rf:cft}
The low energy states of $S$=1/2-TCKM are the NFL states,
and the energies for primary states are known as conformal towers,
\begin{eqnarray}
\label{eq:ECFT}
E_{\rm CFT}(Q_{\rm C}, j, j_f) &=& {{v_{\rm F} \pi} \over \ell}
\left\{ {1 \over 8}
\left(Q_{\rm C} + 2 {\delta_p \over \pi} \right)^2 \right. \nonumber \\
 & & + \left. {j(j+1) \over 4} + {j_f(j_f+1) \over 4} \right\},
\end{eqnarray}
where $Q_{\rm C}, j, j_f$ and $\delta_p$ denote the freedoms
of the charge, the spin and the flavor-spin,
and the potential scattering, respectively.
As seen from the particle hole symmetry of the first excited states
obtained by the NRG calculation, $\delta_p$ becomes zero.
In Table~\ref{table:1} we compare the low energy states
from the NRG calculation and those of the CFT.
The renormalized energy, $E$, in the column of the NRG is
multiplied by a constant factor, $r^*$,
so the energy of the first excited states agrees with that of the CFT.
The correspondence of the charge and the degeneracy 
between both results is perfect.
The difference in energy is small for states in lower energy region
with $r^*E \sim 0.5$,
and it increases for ones in larger energy region, $r^*E \sim 1.0$.
But the difference decreases when the discretization parameter,
$\Lambda$ in the NRG method, is decreased as shown in Fig.~\ref{fig:2}.
It will disappear in the continuum limit, $\Lambda \rightarrow 1$.
This behavior is the same as that of the $S$=1/2-TCKM.~\cite{rf:affleck}
It can be concluded the NFL part of the low energy spectrum in Fig.~\ref{fig:1}
has the same form to that of the $S$=1/2-TCKM.
\begin{fulltable}
\caption{
Comparison between the states belong to the NFL part for the odd NRG step
of the f$^2$ - f$^1$ dominant fluctuation case.
The states at the low energy fixed point from the NRG calculation
and the low energy states for the $S$=1/2-TCKM expected from the CFT are listed.
The parameters for the NRG calculation are the same as those in Fig.~\ref{fig:1}.
The states in the column of NRG are the results at $L=31$.
Index in the column of NRG is the same in Fig.~\ref{fig:1}.
The charge $Q - Q_0$, which is defined from
in Fig.~\ref{fig:1} and $Q_0 = -1$ of the ground state,
coincides with $Q_{\rm C}$.
The degeneracy of each line in the column of CFT is given by $(2j+1)(2j_f+1)$,
and it coincides with the degeneracy given by NRG.
The energy, $E$, is multiplied by a factor $r^*=0.5924$ which is defined so to agree
the energy of the first excited state from the NRG and the CFT results.
In the column of CFT the ground state energy, 3/16, is subtracted.
}
\label{table:1}
\begin{fulltabular}
{@{\hspace{\tabcolsep}\extracolsep{\fill}}
c r c r r c r r r r}
\hline
\multicolumn{5}{c}{NRG}
& \ 
& \multicolumn{4}{c}{CFT} \\
\cline{1-5} \cline{7-10}
\multicolumn{1}{c}{index}
&\multicolumn{1}{c}{$Q-Q_0$}
&\multicolumn{1}{c}{degen.}
&\multicolumn{1}{c}{$E$}
&\multicolumn{1}{c}{$r^*E$}
& \ 
&\multicolumn{1}{c}{$Q_{\rm C}$}
&\multicolumn{1}{c}{$j$}
&\multicolumn{1}{c}{$j_f$}
&\multicolumn{1}{c}{$E_{\rm CFT} \ell / \pi v_{\rm F}$} \\
\hline
1, 2 & 0 & 2 & 0 & 0
& \ 
& 0 & 1/2 & 0 & 0 \\
3, 5 & 1 & 2 & 0.211 & 0.125
& \ 
& 1 & 0 & 1/2 & 1/8 \\
4, 6 & -1 & 2 & 0.211 & 0.125
& \ 
&-1 & 0 & 1/2 & 1/8 \\
11, 12, 14, 17, 19, 20 & 0 & 6 & $0.889 \sim 0.891$ & $0.527 \sim 0.528$
& \ 
& 0 & 1/2 & 1 & 1/2 \\
13, 15 & -2 & 2 & 0.890 & 0.527
& \ 
& -2 & 1/2 & 0 & 1/2 \\
16, 18 & 2 & 2 & 0.890 & 0.527
& \ 
& 2 & 1/2 & 0 & 1/2 \\
33, 34, 39, 40, 43, 44 & 1 & 6 & $1.17 \sim 1.18$ & $0.693 \sim 0.699$
& \
& 1 & 1 & 1/2 & 5/8 \\
35-38, 41, 42 & -1 & 6 & $1.17 \sim 1.18$ & $0.693 \sim 0.699$
& \ 
& 1 & 1 & 1/2 & 5/8 \\
55-58 & 0 & 4 & 1.77 &  1.05
& \ 
& 0 & 3/2 & 0 & 1 \\
79, 80 & 0 & 2 & 1.87 &  1.11
& \ 
& 0 & 1/2 & 0 & 1 \\
113, 114, 117-120 & -2 & 6 & $2.05 \sim 2.06$ & $1.21 \sim 1.22$
& \ 
& -2 & 1/2 & 1 & 1 \\
115, 116 & 0 & 2 & 2.06 &  1.22
& \ 
& 0 & 1/2 & 0 & 1 \\
121-126 & 0 & 6 & 2.06 & 1.22
& \ 
& 0 & 1/2 & 1 & 1 \\
153-158 & 2 & 6 & $2.10 \sim 2.11$ & $1.24 \sim 1.25$
& \ 
& 2 & 1/2 & 1 & 1 \\
\hline
\end{fulltabular}
\end{fulltable}

In the NRG calculation, we have low energy states
which are not listed in Table~\ref{table:1}.
They are explained as the states accompanied by the extra excitations
obeying the LFL theory.
In Table~\ref{table:2} we show the list of these states.
The states, $i=7 \sim 10$ and $i=21 \sim 24$, are interpreted as
the single particle excited state of the LFL part.
The former (latter) are classified to the one hole (electron) excitation
of the occupied (unoccupied) fermion orbitals with energy $-0.884$ (0.988).
From the analysis of the magnetic quantum number the symmetry of the fermion
orbitals of the LFL part is determined as $\Gamma_7$.
The NFL part for these eigen states is in its ground state, $i=1$ and 2.
For the states from $i=25$ to 52 in the Table~\ref{table:2}
the NFL part is in the excited states, $i=3 \sim 6$.
Each state can be explained 
as the product of the NFL state and the LFL state.
One can see the perfect correspondence of the charge
and the magnetic quantum number between the NRG and the expected results.
The energy in the column of LFL is estimated from the sum of
the energies of the NFL state and the LFL part.~\cite{rf:table2}
These energies give good agreement with those of the NRG,
and thus the NFL and LFL parts seem to be almost independent with each other.
The states, $i=53$ and 54, are interpreted as the two hole excited
state of LFL part. The energies of the states agree with the sum of
the energies of two holes as seem from Table~\ref{table:2}.
This is the characteristic features of the LFL excitation.
We have analyzed about 150 states
as the product of the NFL state and the LFL one.
In these state the particle-hole pair excitation
or the two particles excitation of the LFL part are includes.
It is concluded the low energy spectrum at the fixed point
is explained by a model that is the combination of the NFL fixed point
Hamiltonian and the LFL one which are independent with each other.
However, we note the two parts have almost the same low energy scale
because they tend to the low energy fixed point almost simultaneously
at about $L \sim 23$ as seen from Fig.~\ref{fig:1}.
We note the spectrum in the NFL part have the particle-hole symmetry,
but there is the particle-hole asymmetry in the LFL part,
so the total spectrum has the asymmetry.

\begin{table}
\caption{
Comparison between the states belong to the LFL part for odd NRG step
of the f$^2$ - f$^1$ dominant fluctuation case.
The states at the low energy fixed point from the NRG calculation
and the low energy states which is expected by the LFL theory are listed.
The parameters for the NRG calculation are the same as those in Fig.~\ref{fig:1}.
The quantities, $Q, M$ and energy in the column of NRG
is the results for $L=31$ in Fig.~\ref{fig:1}.
All the states in NRG can be explained by the product of
the NFL state and the single particle state which is indicated
in the column of LFL.
The NFL state is indicated by [NFL($i$)], where $i$ is the index of the state
shown in Fig.~\ref{fig:1} and in Table~\ref{table:1}.
The single particle state is indicated by 
$\Gamma_{7 (1)}^+$ and $\Gamma_{7 (1)}^-$, where the plus (minus) sign means
the particle (hole) excitation.
The two hole excited state is indicated by $[\Gamma_{7 (1)}^-]^2$.
The excitation energies, $E(\Gamma_{7 (1)}^+)$ and $E(\Gamma_{7 (1)}^-)$,
is given as 0.988 and 0.844, respectively, to agree with those in the column
of NRG.
The energy in the column of LFL is estimated from the sum of
the NFL state energy and the single particle excitation energy.
}
\label{table:2}
\begin{fulltabular}
{c r r r c l r}
\hline
\multicolumn{4}{c}{NRG}
& \ 
& \multicolumn{2}{c}{LFL} \\
\cline{1-4} \cline{6-7}
\multicolumn{1}{c}{index}
&\multicolumn{1}{c}{$Q$}
&\multicolumn{1}{c}{$2M$}
&\multicolumn{1}{c}{energy}
& \ 
&\multicolumn{1}{c}{state}
&\multicolumn{1}{c}{energy} \\
\hline
7, 9 & -2 & -1, -3 & 0.884
& \ 
& [NFL(1)]$\otimes \Gamma_{7 (1)}^-$ & 0.884 \\
8, 10 & -2 & 1, 3 & 0.884
& \ 
& [NFL(2)]$\otimes \Gamma_{7 (1)}^-$ & 0.884 \\
21, 23 & 0 & -1, -3 & 0.988
& \ 
& [NFL(1)]$\otimes \Gamma_{7 (1)}^+$ & 0.988 \\
22, 24 & 0 & 1, 3 & 0.988
& \ 
& [NFL(2)]$\otimes \Gamma_{7 (1)}^+$ & 0.988 \\
25, 29 & -2 & 0, -2 & 1.096
& \ 
& [NFL(4)]$\otimes \Gamma_{7 (1)}^-$ & 1.095 \\
26, 30 & -3 & 0, 2 & 1.096
& \ 
& [NFL(6)]$\otimes \Gamma_{7 (1)}^-$ & 1.095 \\
27, 31 & -1 & 2, 0 & 1.096
& \ 
& [NFL(3)]$\otimes \Gamma_{7 (1)}^-$ & 1.095 \\
28, 32 & -1 & -2, 0 & 1.096
& \ 
& [NFL(5)]$\otimes \Gamma_{7 (1)}^-$ & 1.095 \\
45, 49 & 1 & 0, 2 & 1.200
& \ 
& [NFL(3)]$\otimes \Gamma_{7 (1)}^+$ & 1.199 \\
46, 50 & 1 & 0, 2 & 1.200
& \ 
& [NFL(5)]$\otimes \Gamma_{7 (1)}^+$ & 1.199 \\
47, 51 & -1 & -2, 0 & 1.201
& \ 
& [NFL(4)]$\otimes \Gamma_{7 (1)}^+$ & 1.199 \\
48, 52 & -1 & 2, 0 & 1.201
& \ 
& [NFL(6)]$\otimes \Gamma_{7 (1)}^+$ & 1.199 \\
53 & -3 & 2 & 1.768
& \ 
& [NFL(1)]$\otimes [\Gamma_{7 (1)}^-]^2$ & 1.768 \\
54 & -3 & -2 & 1.768
& \ 
& [NFL(2)]$\otimes [\Gamma_{7 (1)}^-]^2$ & 1.768 \\
\hline
\end{fulltabular}
\end{table}

In Fig.~\ref{fig:3} we show the FCEL for the even renormalization step.
The parameters are the same as those in Fig.~\ref{fig:1}.
The energy flow is very complicated in the intermediate NRG step,
but, the excitation spectrum at the larger step ($L > 26$)
for the low energy fixed point is simple.
The spectrum is explained by the same model as that
for the odd renormalization step discussed above, that is
the combination of the NFL fixed point model and the LFL one.
The spectrum of the NFL part at the low energy fixed point
has the same form as that for $S$=1/2-TCKM as shown in Table~\ref{table:3}.
The energy, $E$, has the same quantity as that of the odd NRG step.
This is the characteristic behavior of the NFL state.
In Table~\ref{table:4} the low energy states including the excitation of
the LFL part are shown.
The states given by the NRG calculation can be explained by
the combination of the LFL and the NFL states as shown in the column of LFL.
The fermion orbitals have the $\Gamma_7$ symmetry.
Only one orbital above the Fermi level is found
even though we analyze about 150 states.
The single hole excitation is expected to have an energy larger than 2.1.

\begin{fullfigure}
\epsfxsize=.9\linewidth
\epsfbox{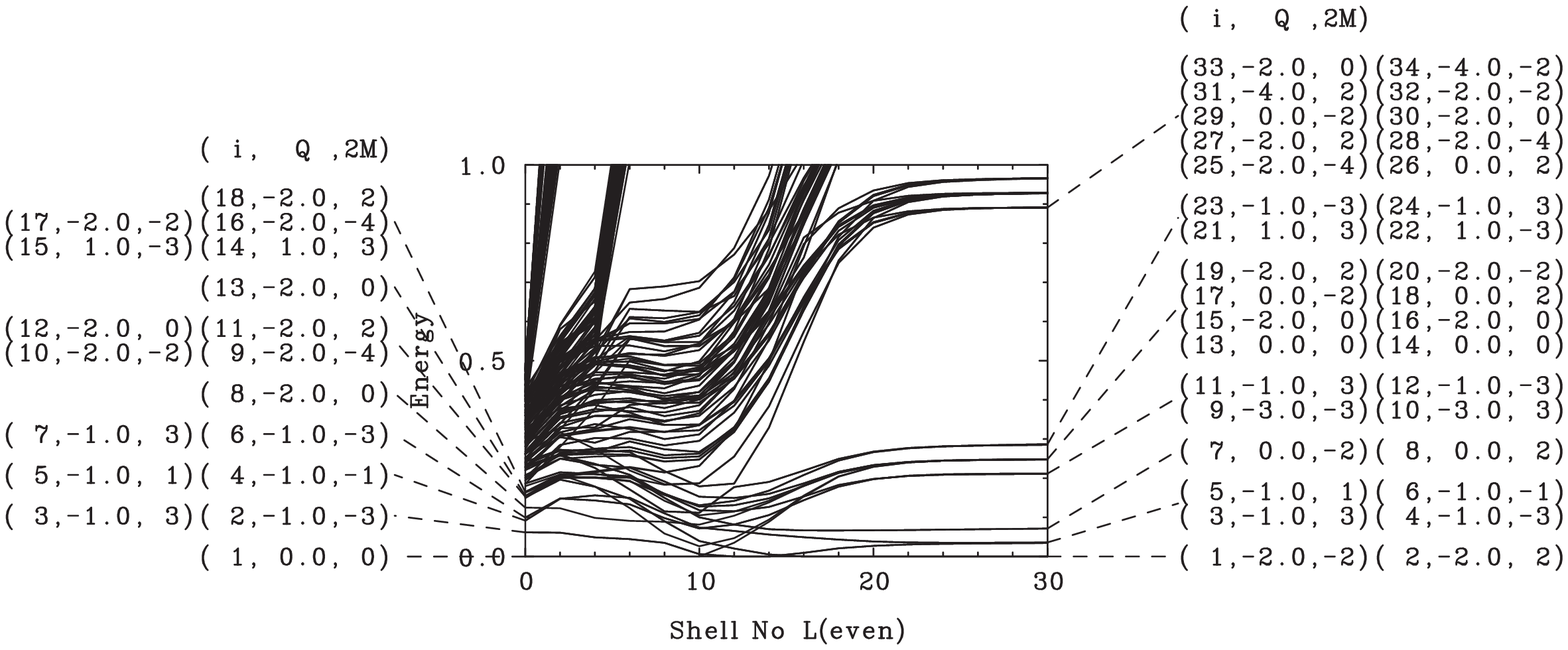}
\caption{
FCEL of the f$^2$ - f$^1$ dominant fluctuation case
for the even renormalization step $L$.
The parameters for the NRG calculation are the same as those in Fig.~\ref{fig:1}.
See the caption of Fig.~\ref{fig:1}.
}
\label{fig:3}
\end{fullfigure}
\begin{fulltable}
\caption{
Comparison between the states belong to the NFL part for even NRG step
of the f$^2$ - f$^1$ dominant fluctuation case.
The states at the low energy fixed point from the NRG calculation
and the low energy states for the $S$=1/2-TCKM expected from the CFT
are listed.
The parameters for the NRG calculation are the same as those in Fig.~\ref{fig:1}.
The states in the column of NRG are the results at $L=30$.
The multiplying factor is $r^*=0.5924$.
In the column of CFT the ground state energy, 3/16, is subtracted.
See the caption of Table~\ref{table:1}.
}
\label{table:3}
\begin{fulltabular}
{c r c r r c r r r r}
\hline
\multicolumn{5}{c}{NRG}
& \ 
& \multicolumn{4}{c}{CFT} \\
\cline{1-5} \cline{7-10}
\multicolumn{1}{c}{index}
&\multicolumn{1}{c}{$Q-Q_0$}
&\multicolumn{1}{c}{degen.}
&\multicolumn{1}{c}{$E$}
&\multicolumn{1}{c}{$r^*E$}
& \ 
&\multicolumn{1}{c}{$Q_{\rm C}$}
&\multicolumn{1}{c}{$j$}
&\multicolumn{1}{c}{$j_f$}
&\multicolumn{1}{c}{$E_{\rm CFT} \ell / \pi v_{\rm F}$} \\
\hline
1, 2 & 0 & 2 & 0 & 0
& \ 
& 0 & 1/2 & 0 & 0 \\
9, 10 & -1 & 2 & 0.211 & 0.125
& \ 
& -1 & 0 & 1/2 & 1/8 \\
11, 12 & 1 & 2 & 0.211 & 0.125
& \ 
& 1 & 0 & 1/2 & 1/8 \\
25, 27, 28, 30, 32, 33 & 0 & 6 & 0.891 & 0.528
& \ 
& 0 & 1/2 & 1 & 1/2 \\
26, 29 & 2 & 2 & 0.891 & 0.528
& \ 
& 2 & 1/2 & 0 & 1/2 \\
31, 34 & -2 & 2 & 0.891 & 0.528
& \ 
& -2 & 1/2 & 0 & 1/2 \\
65-70 & -1 & 6 & $1.17 \sim 1.18$ & $0.693 \sim 0.699$
& \
& -1 & 1 & 1/2 & 5/8 \\
71-76 & 1 & 6 & 1.18 & 0.699
& \ 
& 1 & 1 & 1/2 & 5/8 \\
113-116 & 0 & 4 & 1.77 &  1.05
& \ 
& 0 & 3/2 & 0 & 1 \\
129, 130, 135, 136, 147, 148 & -2 & 6 & $2.10 \sim 2.11$ & $1.24 \sim 1.25$
& \ 
& -2 & 1/2 & 1 & 1 \\
131-134, 145, 146 & 0 & 6 & 2.11 & 1.25
& \ 
& 0 & 1/2 & 1 & 1 \\
137-139, 141, 143, 144 & 2 & 6 & 2.11 & 1.25
& \ 
& 2 & 1/2 & 1 & 1 \\
140, 142 & 0 & 2 & 2.11 &  1.25
& \ 
& 0 & 1/2 & 0 & 1 \\
\hline
\end{fulltabular}
\end{fulltable}
\begin{table}
\caption{
Comparison between the states belong to the LFL part
for even NRG step of the f$^2$ - f$^1$ dominant fluctuation case.
The states at the low energy fixed point from the NRG calculation
and the low energy states which is expected by the LFL theory are listed.
The parameters for the NRG calculation are the same as those in Fig.~\ref{fig:1}.
The quantities in the column of NRG
are the results for $L=30$ in Fig.~\ref{fig:3}.
The NFL state is indicated by [NFL($i$)], where $i$ is the index of the state
shown in Fig.~\ref{fig:3} and in Table~\ref{table:3}.
See the caption of Table~\ref{table:2}.
}
\label{table:4}
\begin{fulltabular}
{c r r r c l r}
\hline
\multicolumn{4}{c}{NRG}
& \ 
& \multicolumn{2}{c}{LFL} \\
\cline{1-4} \cline{6-7}
\multicolumn{1}{c}{index}
&\multicolumn{1}{c}{$Q$}
&\multicolumn{1}{c}{$2M$}
&\multicolumn{1}{c}{energy}
& \ 
&\multicolumn{1}{c}{state}
&\multicolumn{1}{c}{energy} \\
\hline
3, 5 & -1 & 3, 1 & 0.0345
& \ 
& [NFL(1)]$\otimes \Gamma_{7 (1)}^+$ & 0.0345 \\
4, 6 & -1 & -3, -1 & 0.0345
& \ 
& [NFL(2)]$\otimes \Gamma_{7 (1)}^+$ & 0.0345 \\
7 & 0 & -2, 2 & 0.0714
& \ 
& [NFL(1)]$\otimes 2\Gamma_{7 (1)}^+$ & 0.0690 \\
8 & 0 & 2, 2 & 0.0715
& \ 
& [NFL(2)]$\otimes 2\Gamma_{7 (1)}^+$ & 0.0690 \\
13, 17 & 0 & 0, -2 & 0.248
& \ 
& [NFL(11)]$\otimes \Gamma_{7 (1)}^+$ & 0.246\\
14, 18 & 0 & 0, 2 & 0.248
& \ 
& [NFL(12)]$\otimes \Gamma_{7 (1)}^+$ & 0.246 \\
15, 19 & -2 & 0, 2 & 0.248
& \ 
& [NFL(9)]$\otimes \Gamma_{7 (1)}^+$ & 0.246\\
16, 20 & -2 & 0, -2 & 0.248
& \ 
& [NFL(10)]$\otimes \Gamma_{7 (1)}^+$ & 0.246 \\
21 & 1 & 3 & 0.285
& \ 
& [NFL(11)]$\otimes [\Gamma_{7 (1)}^+]^2$ & 0.280 \\
22 & 1 & -3 & 0.285
& \ 
& [NFL(12)]$\otimes [\Gamma_{7 (1)}^+]^2$ & 0.280 \\
23 & -1 & -3 & 0.287
& \ 
& [NFL(9)]$\otimes [\Gamma_{7 (1)}^+]^2$ & 0.280 \\
24 & -1 & 3 & 0.287
& \ 
& [NFL(10)]$\otimes [\Gamma_{7 (1)}^+]^2$ & 0.280 \\
\hline
\end{fulltabular}
\end{table}

There are three kinds of orbitals for the conduction electrons,
$\Gamma_7^{(1)}$, $\Gamma_7^{(2)}$ and $\Gamma_6$.
Two of them will contribute to the NFL state,
and remaining one will be left as the LFL part.
To find which orbitals are related to the NFL state,
we consider a fictitious model that the conduction electrons with
the $\Gamma_7^{(1)}$ symmetry are removed from the original model
given in eq.~(\ref{eq:hamiltonian}).
In Fig.~\ref{fig:4} the FCEL of the fictitious model is shown.
At the low energy fixed point the energy spectrum does not depend on
whether the NRG step is odd or even.
The low energy spectrum is almost the same as that of the NFL part
of the original model.
There are the small particle-hole asymmetry in the spectrum of the fictitious model.
All the states at the low energy fixed point can be analyzed as
the conformal towers of the $S$=1/2-TCKM with the potential scattering term.
The conduction electrons with $\Gamma_6$ and $\Gamma_7^{(2)}$ symmetries
contribute to the NFL state,
and the electrons with $\Gamma_7^{(1)}$ to the LFL state.

When we consider a fictitious model that the conduction electrons with
the $\Gamma_6$ or $\Gamma_7^{(2)}$ symmetry are removed,
the low energy states are explained by a Ising type model:
One localized Ising spin couples with the conduction electrons
through the exchange interaction.

\begin{fullfigure}
\epsfxsize=.9\linewidth
\epsfbox{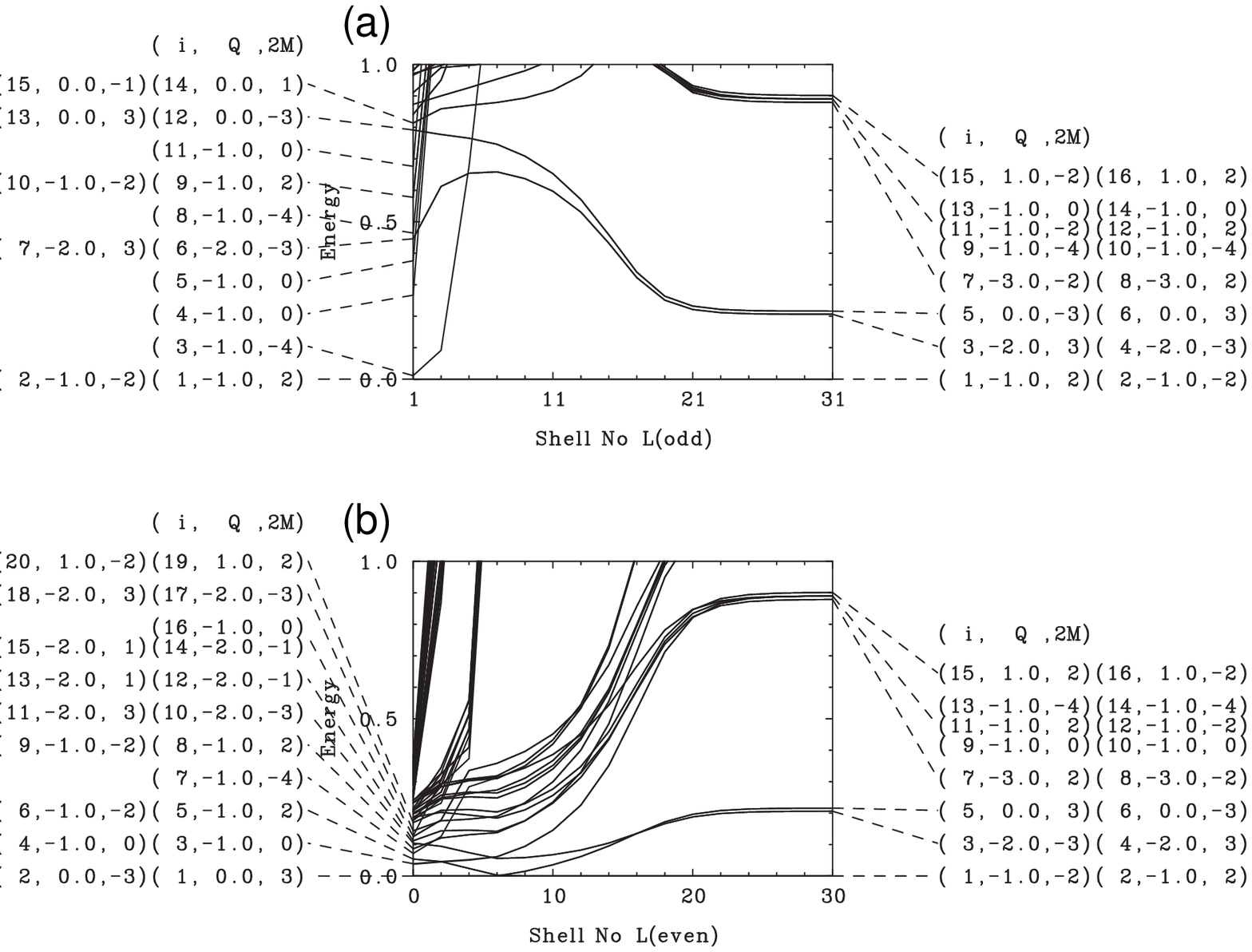}
\caption{
FCEL of the fictitious model for the odd~(a) and even~(b)
renormalization steps $L$.
The parameters for the NRG calculation are the same as those in Fig.~\ref{fig:1}.
See the caption of Fig.~\ref{fig:1}.
}
\label{fig:4}
\end{fullfigure}
\subsection{Flow chart of energy levels of f$^2$ - f$^3$ dominant fluctuation case}
For the f$^2$ - f$^3$ dominant fluctuation case
we choose the parameters;
$\ve(\Gamma_7^{(1)})=-1.2$,
$\ve(\Gamma_6)=-1.05$, $\ve(\Gamma_7^{(2)})=-0.8$, $U=0.3$ and $I=8$.
The f-levels is chosen to be deeper
and the Coulomb interaction to be smaller
than those for the f$^2$ - f$^1$ dominant fluctuation case
with fixing the CEF splitting of f-level and the exchange interaction.
The main states of the f$^2$ configuration are almost the same as those
of the previous case.
The lowest CEF state is the non-Kramers doublet which has the energy, -2.346,
and it has mainly the character of
$J=4$, $\Gamma_5$-irreducible representation of D$_4$-group.
The first to the forth excited states are the singlet
which has mainly the character of 
$\Gamma_4$, $\Gamma_1$, $\Gamma_2$
and $\Gamma_3$-irreducible representation of D$_4$-group, respectively.
The energy of each state is -2.342, -2.186, -2.108 and -2.067, respectively.
The main states of the f$^3$ configuration have mainly the character
of $J=9/2$.
They are the $\Gamma_6$ doublet, the $\Gamma_7$ doublet
and the another $\Gamma_7$ doublet states with energies
-1.958, -1.927 and -1.907, respectively.
When the hybridization matrix $\Gamma=\pi V^2/2$ is changed from
$0.030\pi$ to $0.038\pi$ for $\Lambda=8$,
the occupation number of the f-electron varied
from 2.233 to 2.255.

In Fig.~\ref{fig:5}~(a) the FCEL for the odd NRG step is shown.
Each energy spectrum at the low energy fixed point is
explained by the combination of the NFL fixed point Hamiltonian
and the LFL one which are independent with each other also in this case.
In Table~\ref{table:5}
we show the comparison between the low energy states in the NFL part
of the NRG result and those of the conformal tower of $S$=1/2-TCKM.
The first excited state is the one-particle excitation doublet ($i=3, 4$),
and the second excited state is the one-hole excitation doublet ($i=5, 6$).
There exists the particle-hole asymmetry, so the multiplying factor, $r^*=0.592$,
and the potential scattering, $\delta_p=-0.235$, are determined
from the first and the second excited states.
The states of the NRG result show good correspondence to the result of CFT.
We note the difference of the energies has comparable magnitude
as that found in Table~\ref{table:1}.
It decreases when we use $\Lambda=4$, and it will be removed in the continuum
limit, $\Lambda \rightarrow 1$.
The states accompanied by the LFL excitation are not listed in the table.
The LFL states are explained as the excitations of the orbital
with the $\Gamma_7$ symmetry.
This situation is common to the f$^2$ - f$^1$ dominant fluctuation case.

In Fig.~\ref{fig:5}~(b) we show the FCEL for the even NRG step.
The NFL part of the low energy states are listed in Table~\ref{table:6}.
The quantities, $r^*$ and $\delta_p$, don't depend on
the even or oddness of the NRG step.
The LFL part is not listed in the table.
The first excited states which are indicated by the index, $i=3 \sim 6$,
in the right hand side of the figure are the one hole excitations
of the fermion orbital with $\Gamma_7$ symmetry from the ground state doublet
in the NFL part.
The second excited states are the two holes excitations from the ground state.
As seen from the figure there are the one hole excitations
and the two holes excitations
of the LFL part from each NFL state.
\begin{fullfigure}
\epsfxsize=.9\linewidth
\epsfbox{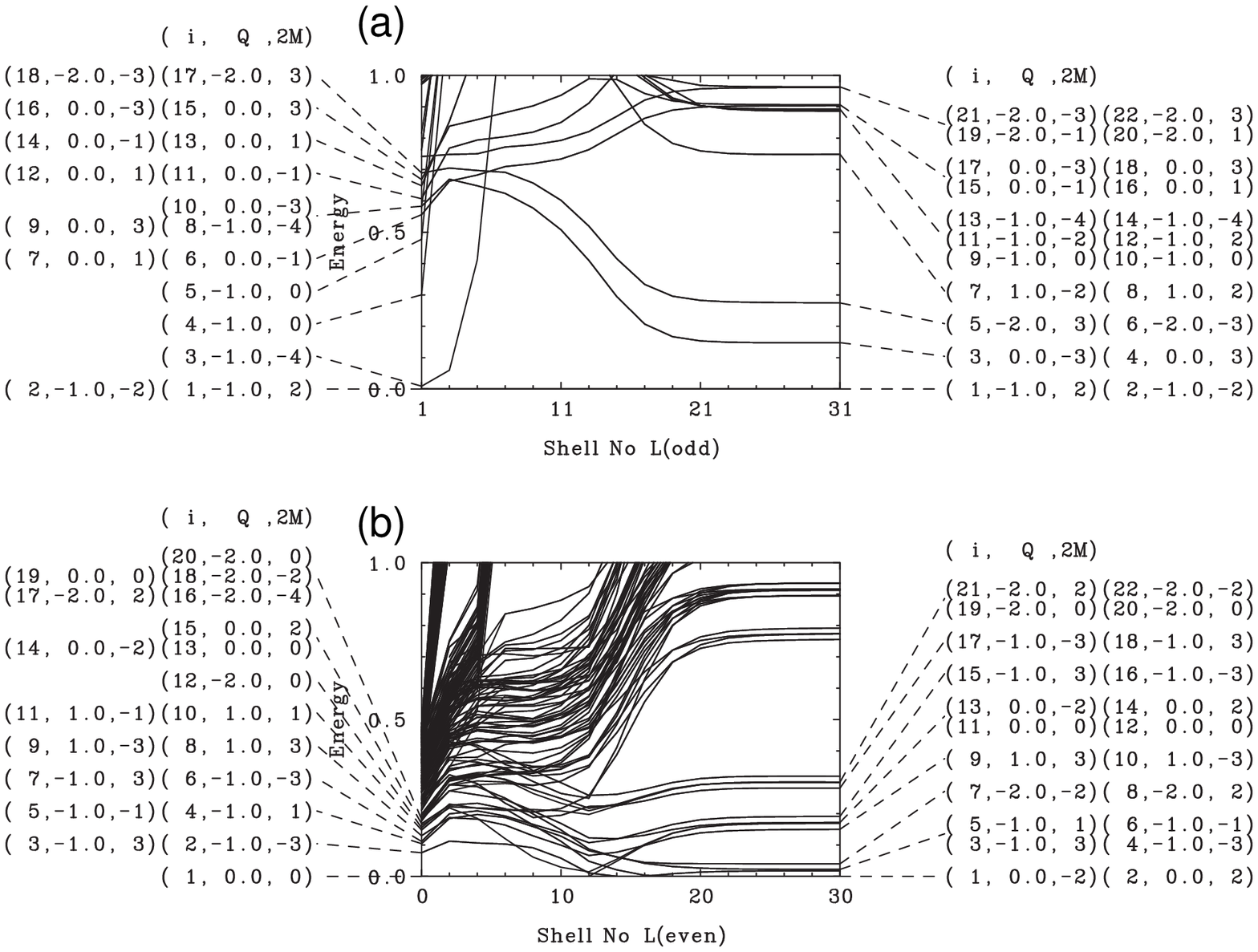}
\caption{
FCEL of the f$^2$ - f$^3$ dominant fluctuation case
for the odd~(a) and the even~(b)
renormalization steps $L$.
The discretization parameter $\Lambda=8$ is used,
and about 500 states are retained at each step.
The parameters of IAM are;
$\ve(\Gamma_7^{(1)})=-1.2$,
$\ve(\Gamma_6)=-1.05$, $\ve(\Gamma_7^{(2)})=-0.8$,
$U=0.3$, $I=8$ and $V^2/2=0.036$.
The occupation number of the f-electron becomes 2.21.
See the caption of Fig.~\ref{fig:1}.
}
\label{fig:5}
\end{fullfigure}
\begin{fulltable}
\caption{
Comparison between the states belong to the NFL part for the odd NRG step
of the f$^2$ - f$^3$ dominant fluctuation case.
The states at the low energy fixed point from the NRG calculation 
and the low energy states for the $S$=1/2-TCKM expected from the CFT are listed.
The parameters for the NRG calculation are the same as those in Fig.~\ref{fig:5}.
The states in the column of NRG are the results at $L=31$.
The multiplying factor, $r^*=0.5920$,
and the potential scattering, $\delta_p=-0.2350$, are defined to agree
the energies of the first and the second excited states
from the NRG and the CFT results.
In the column of CFT the ground state energy, $(\pi \delta_p/2)^2 + 3/16$,
is subtracted.
See the caption of Table~\ref{table:1}.
}
\label{table:5}
\begin{fulltabular}
{@{\hspace{\tabcolsep}\extracolsep{\fill}}
c r c r r c r r r r}
\hline
\multicolumn{5}{c}{NRG}
& \ 
& \multicolumn{4}{c}{CFT} \\
\cline{1-5} \cline{7-10}
\multicolumn{1}{c}{index}
&\multicolumn{1}{c}{$Q-Q_0$}
&\multicolumn{1}{c}{degen.}
&\multicolumn{1}{c}{$E$}
&\multicolumn{1}{c}{$r^*E$}
& \ 
&\multicolumn{1}{c}{$Q_{\rm C}$}
&\multicolumn{1}{c}{$j$}
&\multicolumn{1}{c}{$j_f$}
&\multicolumn{1}{c}{$E_{\rm CFT} \ell / \pi v_{\rm F}$} \\
\hline
1, 2 & 0 & 2 & 0 & 0
& \ 
& 0 & 1/2 & 0 & 0 \\
3, 4 & 1 & 2 & 0.148 & 0.0876
& \ 
& 1 & 0 & 1/2 & 0.0876 \\
5, 6 & -1 & 2 & 0.274 & 0.162
& \ 
&-1 & 0 & 1/2 & 0.162 \\
7, 8 & 2 & 2 & 0.749 & 0.443
& \ 
& 2 & 1/2 & 0 & 0.425 \\
9-14 & 0 & 6 & $0.889 \sim 0.892$ & $0.524 \sim 0.528$
& \ 
& 0 & 1/2 & 1 & 1/2 \\
23, 24 & -2 & 2 & 1.03 & 0.610
& \ 
& -2 & 1/2 & 0 & 0.575 \\
29-34 & 1 & 6 & $1.09 \sim 1.10$ & $0.645 \sim 0.648$
& \
& 1 & 1 & 1/2 & 0.588 \\
47-52 & -1 & 6 & $1.26 \sim 1.27$ & $0.748 \sim 0.749$
& \ 
& -1 & 1 & 1/2 & 0.662 \\
\hline
\end{fulltabular}
\end{fulltable}
\begin{fulltable}
\caption{
Comparison between the states belong to the NFL part for the even NRG step
of the f$^2$ - f$^3$ dominant fluctuation case.
The states at the low energy fixed point from the NRG calculation 
and the low energy states for the $S$=1/2-TCKM expected from the CFT are listed.
The parameters for the NRG calculation are the same as those in Fig.~\ref{fig:5}.
The states in the column of NRG are the results at $L=30$.
The multiplying factor and the potential scattering are:
$r^*=0.5799$ and $\delta_p=-0.2393$, respectively.
In the column of CFT the ground state energy, $(\pi \delta_p/2)^2 + 3/16$,
is subtracted.
See the caption of Table~\ref{table:1}.
}
\label{table:6}
\begin{fulltabular}
{@{\hspace{\tabcolsep}\extracolsep{\fill}}
c r c r r c r r r r}
\hline
\multicolumn{5}{c}{NRG}
& \ 
& \multicolumn{4}{c}{CFT} \\
\cline{1-5} \cline{7-10}
\multicolumn{1}{c}{index}
&\multicolumn{1}{c}{$Q-Q_0$}
&\multicolumn{1}{c}{degen.}
&\multicolumn{1}{c}{$E$}
&\multicolumn{1}{c}{$r^*E$}
& \ 
&\multicolumn{1}{c}{$Q_{\rm C}$}
&\multicolumn{1}{c}{$j$}
&\multicolumn{1}{c}{$j_f$}
&\multicolumn{1}{c}{$E_{\rm CFT} \ell / \pi v_{\rm F}$} \\
\hline
1, 2 & 0 & 2 & 0 & 0
& \ 
& 0 & 1/2 & 0 & 0 \\
9, 10 & 1 & 2 & 0.150 & 0.0869
& \ 
& 1 & 0 & 1/2 & 0.0869 \\
17, 18 & -1 & 2 & 0.281 & 0.163
& \ 
&-1 & 0 & 1/2 & 0.163 \\
25, 26 & 2 & 2 & 0.755 & 0.447
& \ 
& 2 & 1/2 & 0 & 0.424 \\
33-38 & 0 & 6 & $0.894 \sim 0.896$ & $0.529 \sim 0.530$
& \ 
& 0 & 1/2 & 1 & 1/2 \\
57, 58 & -2 & 2 & 1.04 & 0.614
& \ 
& -2 & 1/2 & 0 & 0.576 \\
65-70 & 1 & 6 & $1.10 \sim 1.10$ & 0.649
& \
& 1 & 1 & 1/2 & 0.587 \\
89-94 & -1 & 6 & 1.27 & 0.750
& \ 
& -1 & 1 & 1/2 & 0.663 \\
\hline
\end{fulltabular}
\end{fulltable}

When we consider the fictitious model that the conduction electrons
with $\Gamma_7^{(1)}$ symmetry are removed,
the spectrum at the low energy shows a similar NFL behavior of the original model.
In other cases; the fictitious model removing the conduction electrons
with the $\Gamma_6$ or $\Gamma_7^{(2)}$ symmetry,
we have the Ising type fixed point.
Therefore the $\Gamma_6$ and $\Gamma_7^{(2)}$ components contribute
to the NFL part and the $\Gamma_7^{(1)}$ to the LFL part.
This fact is the same as that of the f$^2$ - f$^1$ dominant fluctuation case.

We note when the original model with the large hybridization strength
is considered,
the low energy states are explained by the LFL theory.
The lowest energy state at each NRG step changes
from the crystalline field doublet ($L=-1$) to the singlet state ($L \geq 15$)
in the FCEL of the odd step.
The two electrons are bounded at the intermediate step, $L=15$,
so the singlet state has the character similar to that of the f$^4$ singlet state.
This contrasts with the f$^0$-singlet-like ground state
in the f$^2$ - f$^1$ dominant fluctuation case
with the large hybridization strength.

From the analysis of this section we can deduce the following conclusions:
The low energy fixed point properties of the present model can be given
by the combination of the two independent parts, one is the NFL component
and the another is the LFL component.
The two parts have almost the same low energy scale, and in addition
the separation of these occurs at almost the same energy region
with the low energy scale.
The conduction electrons with $\Gamma_6$ and $\Gamma_7^{(2)}$ symmetry
contribute to the NFL part and the $\Gamma_7^{(1)}$ symmetry to the LFL part.
The above facts are seen commonly both
for the f$^2$ - f$^1$ and the f$^2$ - f$^3$ fluctuation cases.

\section{Magnetization}
We calculate the temperature dependence of the magnetization of f-electrons,
$M$ = $\langle m_f \rangle$ = $\langle \sum_{m=-j}^{j} m f^\dagger_m f_m \rangle$,
by adding the Zeeman term, ${\cal H'}$ = $-m_f H_z$, to eq.~(\ref{eq:hamiltonian}),
where $m$ is the magnetic quantum number of $j$.
The thermal average at the temperature, $T_L$, is calculated
by the eigen states of the $L$th cluster as,
\begin{eqnarray}
M(T_L) &=& {{{\rm Tr} m_f {\rm exp} \left\{
- \left( {\cal H}_L + {\cal H'} \right) / T_L \right\}}
\over
{{\rm Tr} {\rm exp} \left\{
- \left( {\cal H}_L + {\cal H'} \right) / T_L \right\}}}, \\
T_L &=& {1 \over 2} \left( 1+ \Lambda^{-1} \right)
\Lambda^{(L-1)/2} / \bar{\beta},
\end{eqnarray}
where $\bar{\beta} \sim 2 $ is a parameter for suiting the temperature to
the eigen energies of each NRG step.

\begin{figure}
%
\epsfxsize=.4\linewidth
\epsfbox{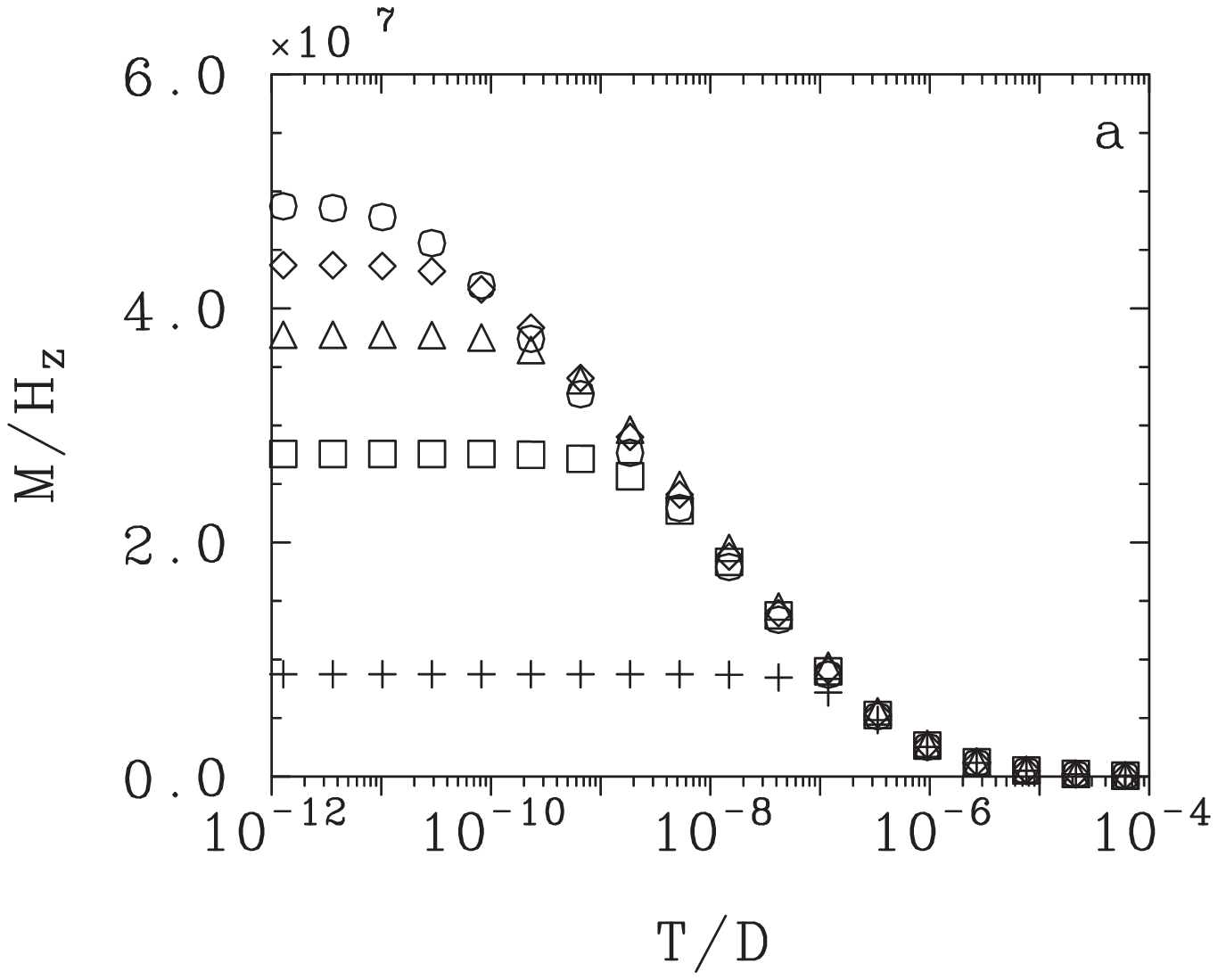}

\epsfxsize=.4\linewidth
\epsfbox{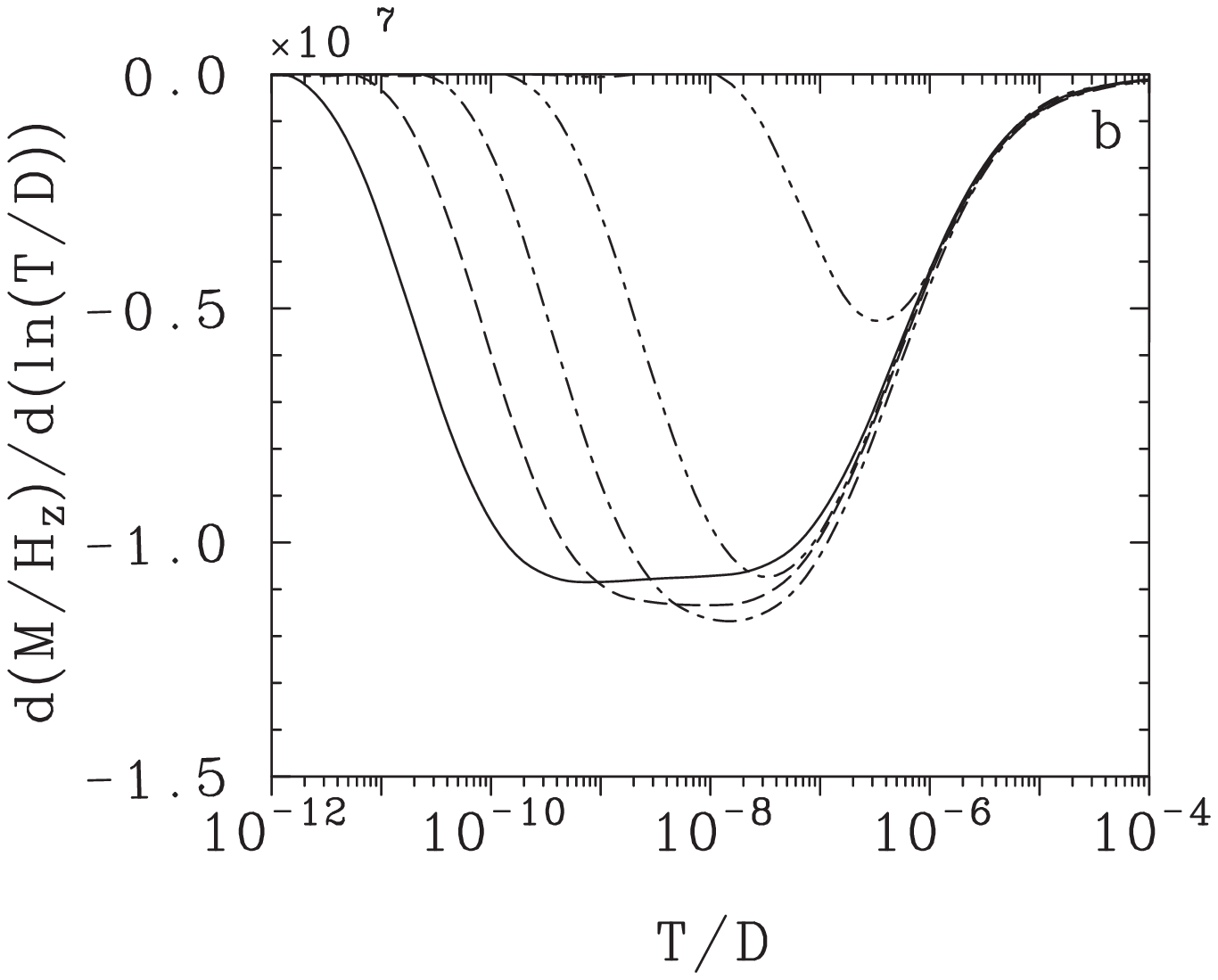}
\caption{
Temperature dependence of the normalized magnetizations~(a)
and the differential of them
by the logarithm of the temperature~(b)
for several magnetic fields,
$H_z$ = $1 \times 10^{-9} (\circ, \mbox{solid line})$,
$2 \times 10^{-9} (\diamond, \mbox{dashed line})$,
$4 \times 10^{-9} (\triangle, \mbox{dot-dashed line})$,
$1 \times 10^{-8} (\Box, \mbox{two-dots-dashed line})$,
and $1 \times 10^{-7} (+, \mbox{three-dots-dashed line})$, respectively.
The parameters are chosen so that
the valence fluctuation between the f$^2$ and f$^3$ configurations becomes
dominant:
$\ve(\Gamma_7^{(1)})=-1.2$,
$\ve(\Gamma_6)=-1.05$, $\ve(\Gamma_7^{(2)})=-0.8$,
$U=0.3$, $I=8$ and $V^2/2=0.035$.
The occupation number of the f-electron for $H_z$ = $1 \times 10^{-9}$ is 2.23.
$\Lambda=8$ is used, and about 400 states are retained at each renormalization step.
}
\label{fig:6}
\end{figure}

In Fig.~\ref{fig:6}~(a) we show the magnetizations
which are normalized by the applied magnetic fields.
These quantities correspond to the magnetic susceptibility when
the field is weak enough.
The parameters of the IAM are chosen so that
the valence fluctuation between the f$^2$ and f$^3$ configurations becomes
dominant.
We use $\Lambda=8$ and about 400 states are kept at each NRG step.
Even when we keep about 500 states, the results of the magnetization are
not changed essentially.
The normalized magnetizations have almost same value,
when the temperature is higher than the Zeeman energy.
For very weak magnetic field
there is a temperature region that the magnetization follows the $- \ln T$
dependence.
When the temperature decreases, the magnetization saturates.
However, $- \ln T$ temperature region extends
with decreasing the applied magnetic field.
This behavior can be seen clearly from Fig.~\ref{fig:6}~(b) which shows
the differential of the magnetizations by the logarithm of the temperature.
The magnetic susceptibility follows $- \ln T$ dependence at low temperatures.
This temperature dependence is the characteristic behavior of the NFL state
for $S$=1/2-TCKM type.
The inverse of the coefficient of $-\ln T$ term is estimated to be
about $10^{-7}$. This means that the energy scale which characterizes
the strength of the $-\ln T$ term of the susceptibility is about $10^{-7}$,
and is very small. Therefore the present model gives sizable $-\ln T$ term.

The saturation of the magnetization relates to a break of the NFL state
under the magnetic field as follows:
In Fig.~\ref{fig:7} the FCEL for the odd NRG step under the magnetic field,
$H_z=1.0 \times 10^{-9}$, is shown.
The energy flow in the region below $L=15$ is almost identical to that
in Fig.~\ref{fig:5}(a)
which is the FCEL without the magnetic field.
The spectrum for $L=15 \sim 21$ splits gradually due to the magnetic field.
However, the behavior is very similar to
that of the NFL spectrum at the low energy fixed point in Fig.~\ref{fig:5}~(a).
The hopping energy at $L=21$, $t_{L-1} \sim \Lambda^{-10} \sim 10^{-9}$,
has the same order to that of the Zeeman energy.
We note that the magnetization shows the saturation deviated from
the $-\ln T$ behavior at about $T \sim 10^{-10}$ for the case of $H_z = 10^{-9}$.
The flow in the region above $L=21$ goes to the LFL spectrum
at the low energy fixed point.
The first and second excited states, $i$=2 and 3,
where $i$ is the index in the figure, are the one electron excitations,
respectively.
The third ($i$=4) and fourth ($i$=5) excited states are the one hole
excitations.
The excited state, $i$=6, is the two electrons excitation which
is explained by the combination of the excitations for the states, $i$=2 and 3.
The excited states, $i=7 \sim 9$, are explained by
the electron-hole pair excitations, and other high energy states are also
explained by the LFL fixed point model.
The NFL state breaks due to the magnetic field, and the low energy states
can be explained by the LFL theory.

\begin{fullfigure}
\epsfxsize=.9\linewidth
\epsfbox{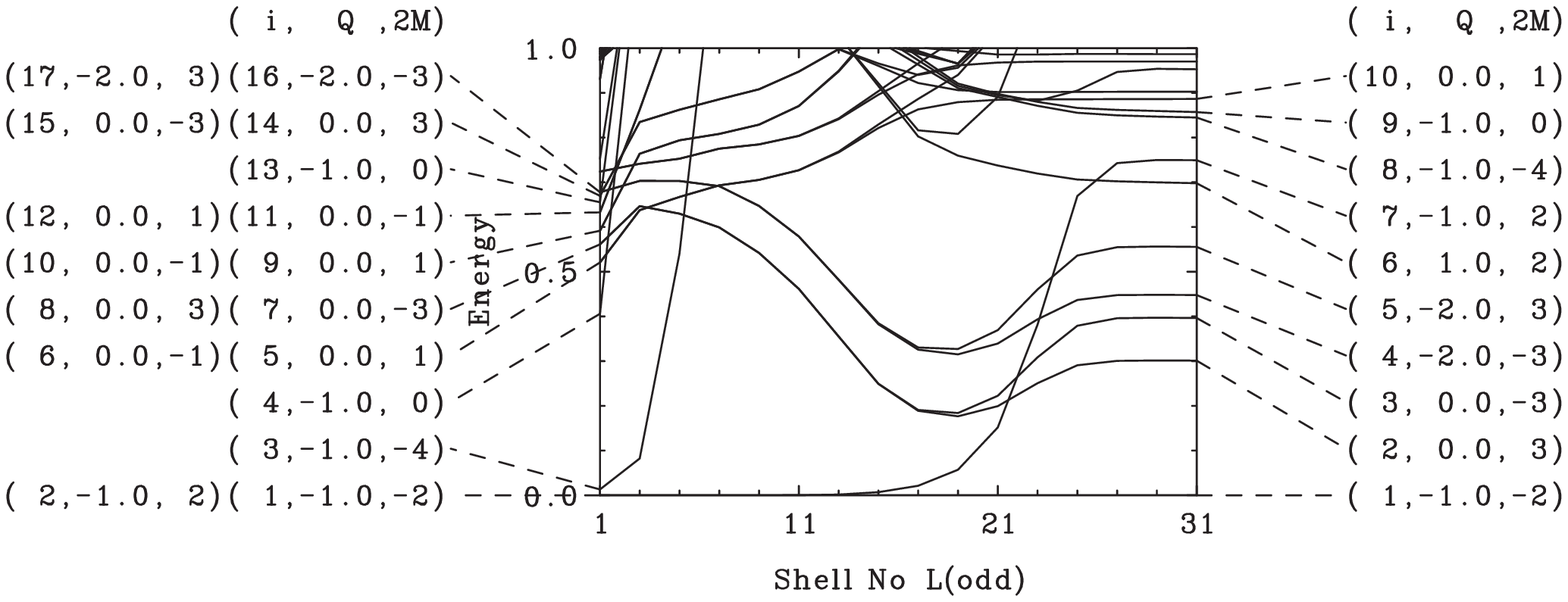}
\caption{
FCEL under the magnetic field, $H_z = 1 \times 10^{-9}$
for the odd renormalization steps $L$.
The parameters are the same as those in Fig.~\ref{fig:6}.
See the caption of Fig.~\ref{fig:1}.
}
\label{fig:7}
\end{fullfigure}

\begin{figure}
\epsfxsize=.4\linewidth
\epsfbox{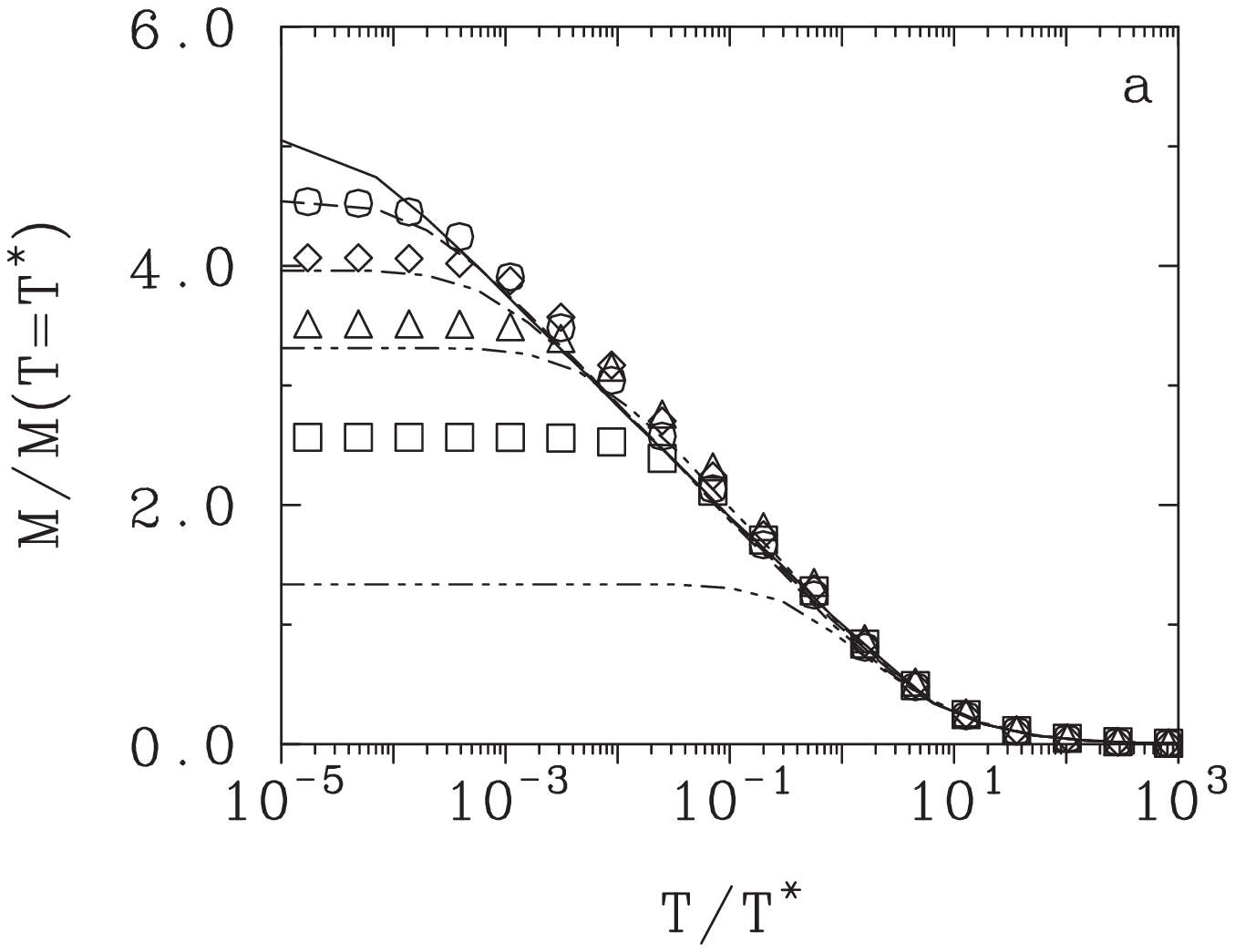}
\caption{
Temperature dependence of scaled magnetizations
for several magnetic fields.
$\Lambda = 8$ is used, and about 400 states are kept in each renormalization step.
The parameters for the magnetizations, which are indicated by the symbols, are
$\ve(\Gamma_7^{(1)})=-1.2$,
$\ve(\Gamma_6)=-1.05$, $\ve(\Gamma_7^{(2)})=-0.8$,
$U=0.3$, $I=8$ and $V^2/2=0.035$, and
$H_z / T^*$ = $1.35 \times 10^{-2} (\circ)$,
$2.7 \times 10^{-2} (\diamond)$,
$5.4 \times 10^{-2} (\triangle)$ and
$0.135 (\Box)$, where $T^*$ = $7.42 \times 10^{-8}$.
The occupation number of the f-electron
for $H_z / T^*$ = $1.35 \times 10^{-2}$ is 2.23.
The magnetizations are normalized by $M^* \equiv M(T=T^*, H_z =1 \times 10^{-9})$
= $1.07 \times 10^{-3}$.
The parameters for the magnetizations which are indicated by the lines are
$\ve(\Gamma_7^{(1)})=-0.9$,
$\ve(\Gamma_6)=-0.75$, $\ve(\Gamma_7^{(2)})=-0.5$,
$U=0.6$, $I=8$ and $V^2/2=0.035$, and
$H_z / T^*$ = $5.5 \times 10^{-3}$ (solid line),
$1.1 \times 10^{-2}$ (dashed line),
$2.2 \times 10^{-2}$ (dot-dashed line),
$5.5 \times 10^{-2}$ (two-dot-dashed line) and
$0.55$ (three-dot-dashed line), where $T^*$ = $1.82 \times 10^{-8}$.
The occupation number of the f-electron
for $H_z / T^*$ = $5.5 \times 10^{-3}$ is 1.91.
$M^* \equiv M(T=T^*, H_z =1 \times 10^{-10})$
= $4.74 \times 10^{-3}$.
}
\label{fig:8}
\end{figure}

In Fig.~\ref{fig:8} we show the magnetizations for the NFL states of various
parameter cases.
It is normalized by
a quantity, $M^*$, where $M^*$ is the magnetization 
under very weak magnetic field at a temperature, $T^*$.
The quantity, $T^*$ is defined as the temperature that $-\ln T$ dependence
of the magnetization begins.
The strength of the $-\ln T$ term is approximately characterized by $1/T^*$.
The symbols shows the magnetizations
for the f$^2$ - f$^3$ dominant fluctuation case,
and the lines for the f$^2$ - f$^1$ dominant fluctuation one.
As seen from the figure the normalized magnetizations under the same
normalized magnetic field, $H_z / T^*$ have almost the same temperature
dependence.
When the magnetic field is very weak, the temperature region
that the magnetization has the $-\ln T$ dependence
spreads from $T^*$ to $H_z /10$.
The model of the f$^2$ - f$^1$ dominant fluctuation also
gives sizable $-\ln T$ divergence.

\section{Magneto Resistance}
In the previous sections we have shown that the low energy spectrum of the present
model is formed by the combination of the NFL and the LFL parts.
We note that the phase shift which is calculated from the spectra
of LFL part becomes very small at low energy fixed point.
Therefore the electric resistivity caused by the LFL part will
decrease with decreasing temperature.
In this section we study the electric resistivity due to the NFL part.
To simplify the analysis
we use the fictitious model which is introduced in \S 3.2.
When the magnetic field is applied, the spectrum at the low energy fixed point
follows the LFL theory.
The single particle excitations are given by the following effective
Hamiltonian,
\begin{eqnarray}
{\cal H}^{\rm eff} &=& {\cal H}_L^0
+ \sum_{\Gamma \gamma} \delta_f^{\rm eff}\left(\Gamma \gamma \right)
f^\dagger_{\Gamma \gamma}f_{\Gamma \gamma} \nonumber \\
&+& \sum_{\Gamma \gamma}
\sqrt{2\Gamma_{\Lambda}^{\rm eff}\left(\Gamma \gamma \right)/ \pi}
\left(f^\dagger_{\Gamma \gamma}s_{0 \Gamma \gamma} + {\rm h.c.}\right),
\end{eqnarray}
where $\delta_f^{\rm eff}\left(\Gamma \gamma \right)$ and
$\Gamma_\Lambda^{\rm eff}\left(\Gamma \gamma \right)$ are the effective
single particle level and effective hybridization width, respectively.~\cite{rf:nrg}
The effective quantities depend on the $\Gamma$ and the suffix $\gamma$
specifying its component, and are determined
by using the low energy excited states.
The phase shift is given as,
\begin{equation}
\eta(\Gamma \gamma) = {\pi \over 2}
+ \arctan{-{\delta_f^{\rm eff}\left(\Gamma \gamma \right)} \over
{\Gamma_\Lambda^{\rm eff}\left(\Gamma \gamma \right) / A_\Lambda}
}.
\end{equation}

\begin{figure}
\epsfxsize=.4\linewidth
\epsfbox{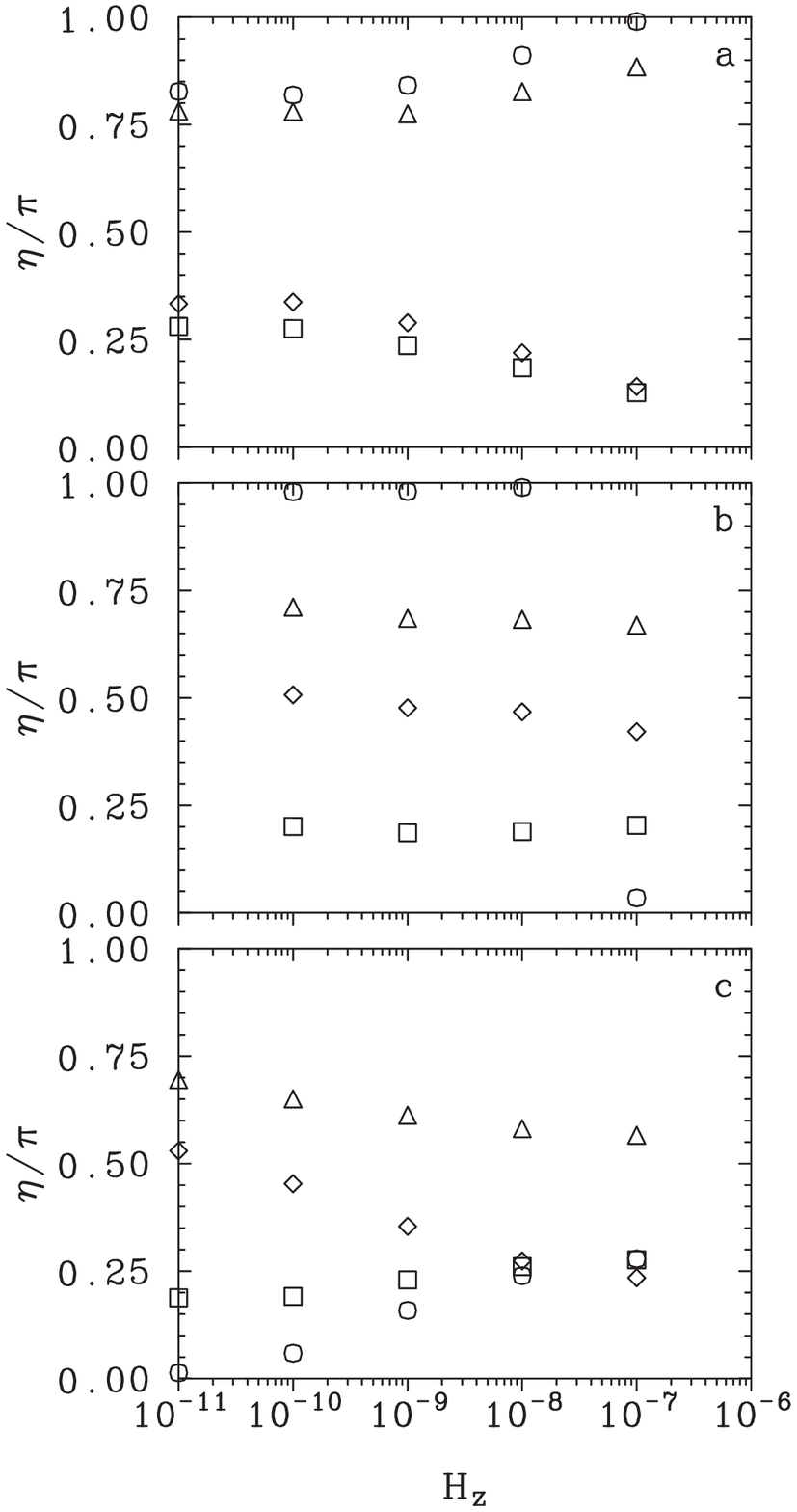}
\caption{Magnetic field dependence of the phase shift.
The parameters are $\ve(\Gamma_7^{(1)})=-1.2$,
$\ve(\Gamma_6)=-1.05$, $\ve(\Gamma_7^{(2)})=-0.8$,
$U=0.3$, $I=8$ and $V^2/2=0.035$~(a), $0.13$~(b) and $0.16$~(c).
The occupation number of the f-electron for $H_z = 0$
are 2.11(a), 2.27(b) and 2.29(c), respectively.
Each symbol shows the phase shift for each component
of the $\Gamma$-irreducible representation;
$[\Gamma_7, +]~(\circ)$, [$\Gamma_7, -]~(\diamond)$,
$[\Gamma_6, +]~(\triangle)$ and $[\Gamma_6, -]~(\Box)$.
}
\label{fig:9}
\end{figure}

Figure~\ref{fig:9} shows the magnetic fields dependence of the phase shift.
The hybridization strength is increased from Figs.~(a) to (c).
We note even though the hybridization is large in Fig.~(c),
the low energy spectrum follows the NFL model for the case of no magnetic field.
Each symbol shows the phase shift for each component
of the $\Gamma$-irreducible representation.
When the hybridization is weak, phase shift has the values about $\pi/4$
and $3\pi/4$ at weak magnetic field limit.
This behavior is same to that of $S$=1/2-TCKM.~\cite{rf:affleck}
When the magnetic field is increased, the phase shift changes gradually,
and the strength of the scattering amplitude for all components decreases.
In the case of the intermediate hybridization, the phase shift scarcely
depends on the magnetic field.
When the magnetic field is increased in the case of the large hybridization,
the phase shift of three components changes to increase
the strength of the scattering amplitude,
while that of one component changes to decrease it.
\begin{figure}
\epsfxsize=.4\linewidth
\epsfbox{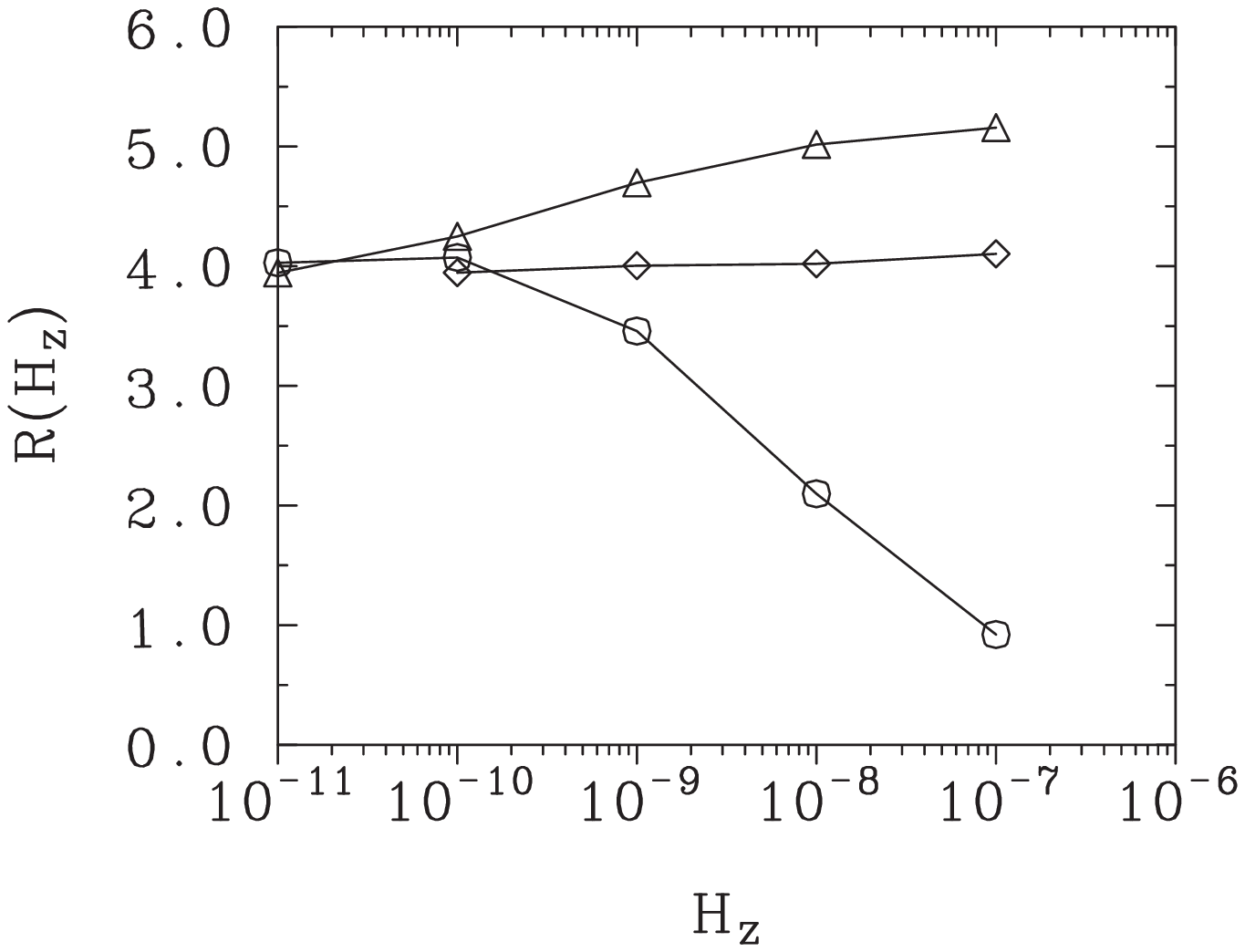}
\caption{Magneto resistance for the several cases of hybridization strength.
The parameters are the same as those in Fig.~\ref{fig:9}~a~($\circ$),
b~($\diamond$) and c~($\triangle$).
}
\label{fig:10}
\end{figure}
\begin{figure}
\epsfxsize=.4\linewidth
\epsfbox{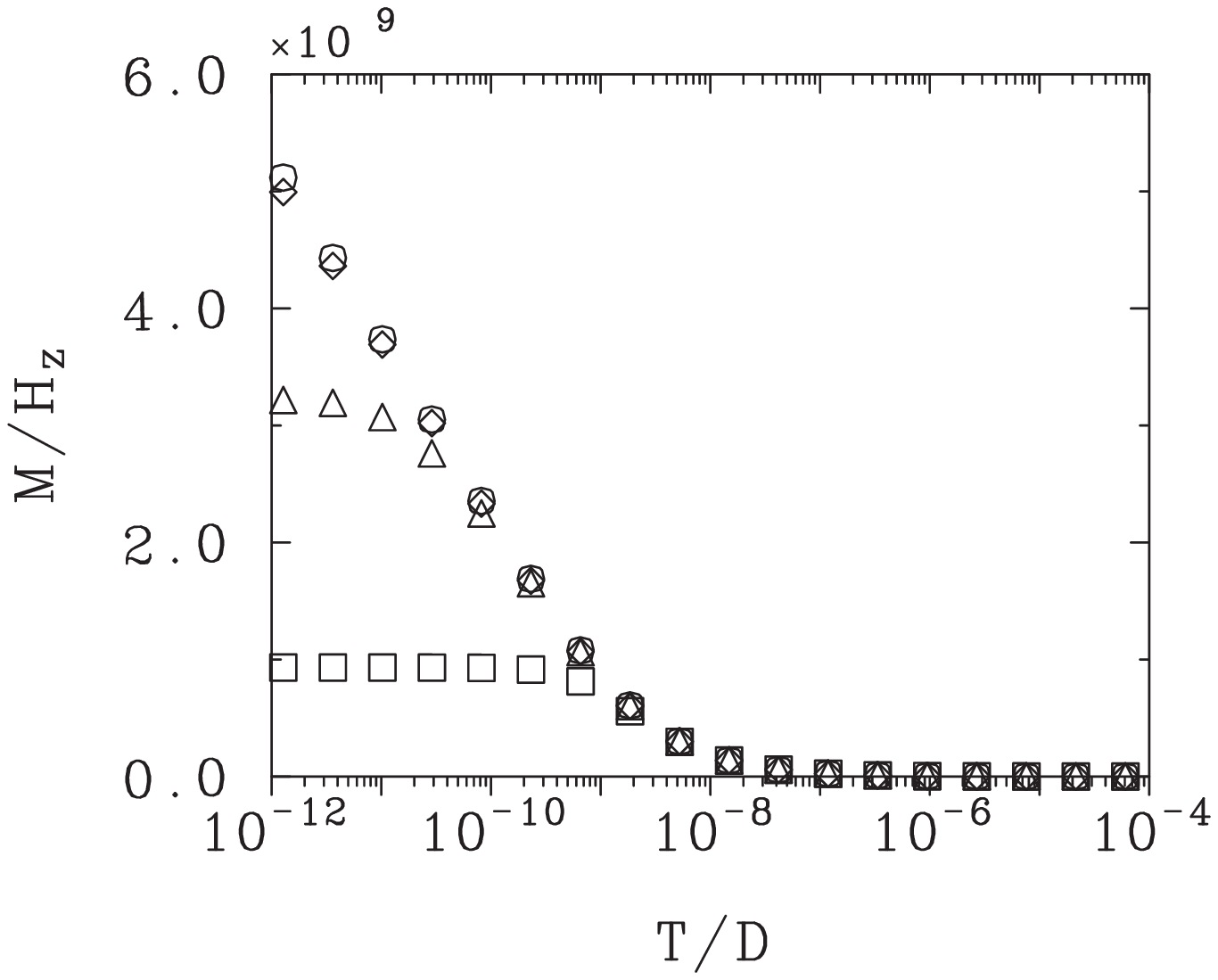}
\caption{Temperature dependence of the normalized magnetizations
for the positive magneto resistance case.
The parameters are the same as those in Fig.~\ref{fig:9}~(c).
The magnetic fields are
$H_z$ = $1 \times 10^{-12} (\circ)$,
$2 \times 10^{-11} (\diamond)$,
$4 \times 10^{-10} (\triangle)$ and
$1 \times 10^{-9} (\Box)$, respectively.
}
\label{fig:11}
\end{figure}

We calculate the magneto resistance by assuming that four components
contribute with equal weight,
\begin{equation}
R(H_z) = \sum_{\Gamma \gamma}
\left( 1 - \cos {2\eta \left(\Gamma \gamma \right)} \right).
\end{equation}
Figure~\ref{fig:10} shows the magneto resistance at $T=0$,
where the circle, diamond and triangle are calculated from the phase shift
shown in Figs.~\ref{fig:9}(a), (b) and (c), respectively.
Each resistance goes to 4.0 in the weak magnetic field limit.
This value is half of the unitarity limit value.
When the hybridization is small, we have the negative magneto resistance,
while we have the positive magneto resistance
for the large hybridization case.
The magnetization for the case of large hybridization is shown in Fig.~\ref{fig:11}.
We can see the $-\ln T$ dependence with the large coefficient.
The coefficient of the $-\ln T$ term is large also for the small hybridization case.
When the hybridization is intermediate, 
the resistance hardly depends on the magnetic field,
and the temperature, $T^*$, becomes high.
This indicates that the initial; namely, the non-renormalized model is
close to the low energy fixed point model.
We note that the coefficient of the $-\ln T$ term is small
in this case.

We note that the positive magneto resistance
is recently observed at low temperature
for U$_x$Th$_{1-x}$Ru$_2$Si$_2$.~\cite{rf:ami4}
For the positive magneto resistance case
we can expect that the electric resistivity decreases
with decreasing temperature.
In fact, Affleck~{\it et al.} have demonstrated that
the magneto resistance is positive for the $S$=1/2-TCKM with very strong
exchange coupling case.~\cite{rf:affleck}
In this case we can expect that the resistivity decreases with decreasing temperature.
The renormalized coupling decreases with decreasing temperature for the strong
coupling model because the low energy fixed point of the NFL state
is characterized by the intermediate coupling value.
We can also expect the positive magneto resistance when the low energy
fixed point of the NFL state is approached from the stronger coupling side with
renormalization step.
The larger hybridization case of the present model seems to show this situation.
We note that the present model can give the positive magneto resistance
together with the sizable $-\ln T$ term of the magnetic susceptibility.
However, the $S$=1/2-TCKM has very small coefficient of the $-\ln T$ term
with magnitude, $1/D$, because the exchange constant must have comparable
value with the band width to get the positive magneto resistance.

From the calculations shown in this section we can expect that the anomalous
properties of U$_x$Th$_{1-x}$Ru$_2$Si$_2$ can be explained within the scenario
of the NFL anomaly of the TCKM type.
Hear we note the temperature dependence of the resistivity of the ETCAM.
The resistivity of the ETCAM
decreases with decreasing temperature, when the large hybridization
is assumed.~\cite{rf:ours3}
In the very low energy region of the NFL state,
the scattering amplitude decreases to half of the unitarity limit value
from the lager value (near the unitarity limit) of the preceding energy region.
The preceding region is characterized by the Kondo effect
of the two-channel Anderson model with larger hybridization.
The behavior of the FCEL of the ETCAM have similar
characteristics to that of the present model with larger hybridization.
It shows the temperature dependence of
the magnetization and the resistivity similar to that of U$_x$Th$_{1-x}$Ru$_2$Si$_2$.
The ETCAM seems to be an effective model of the NFL part of the realistic model
at low temperature.
The local spin in the ETCAM will correspond to
the f-electron with $\Gamma_7^{(1)}$ symmetry,
while the two channels will correspond to electrons
with $\Gamma_7^{(2)}$ and $\Gamma_6$ symmetries of the present model.
However, at present it is not easy to relate the interaction terms
between two models.

\section{Stability of non-Fermi-liquid State}
In the above sections we have shown the NFL behavior of the system
with the non-Kramers doublet CEF ground state.
Because the low energy spectrum of the NFL part has the same form as that
of $S$=1/2-TCKM, an entropy at the zero temperature is expected to remain.
However, it conflicts with the law of thermodynamics.
The residual entropy should be released by some remaining mechanisms.
We have shown when the magnetic field is applied, the NFL state breaks
below the low energy, $H_z/10$, so the residual entropy is released.
\begin{figure}
%
\epsfxsize=.4\linewidth
\epsfbox{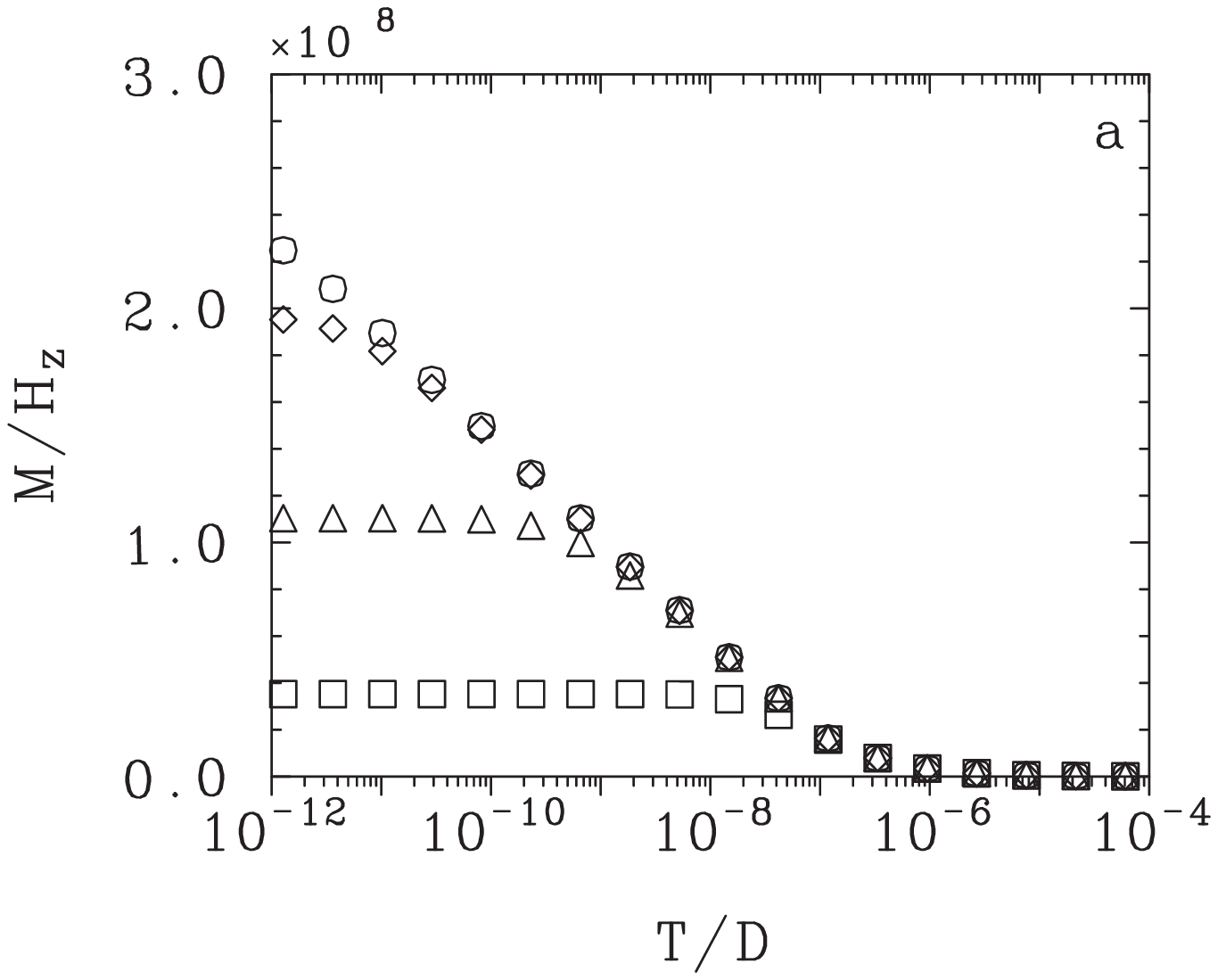}

\epsfxsize=.4\linewidth
\epsfbox{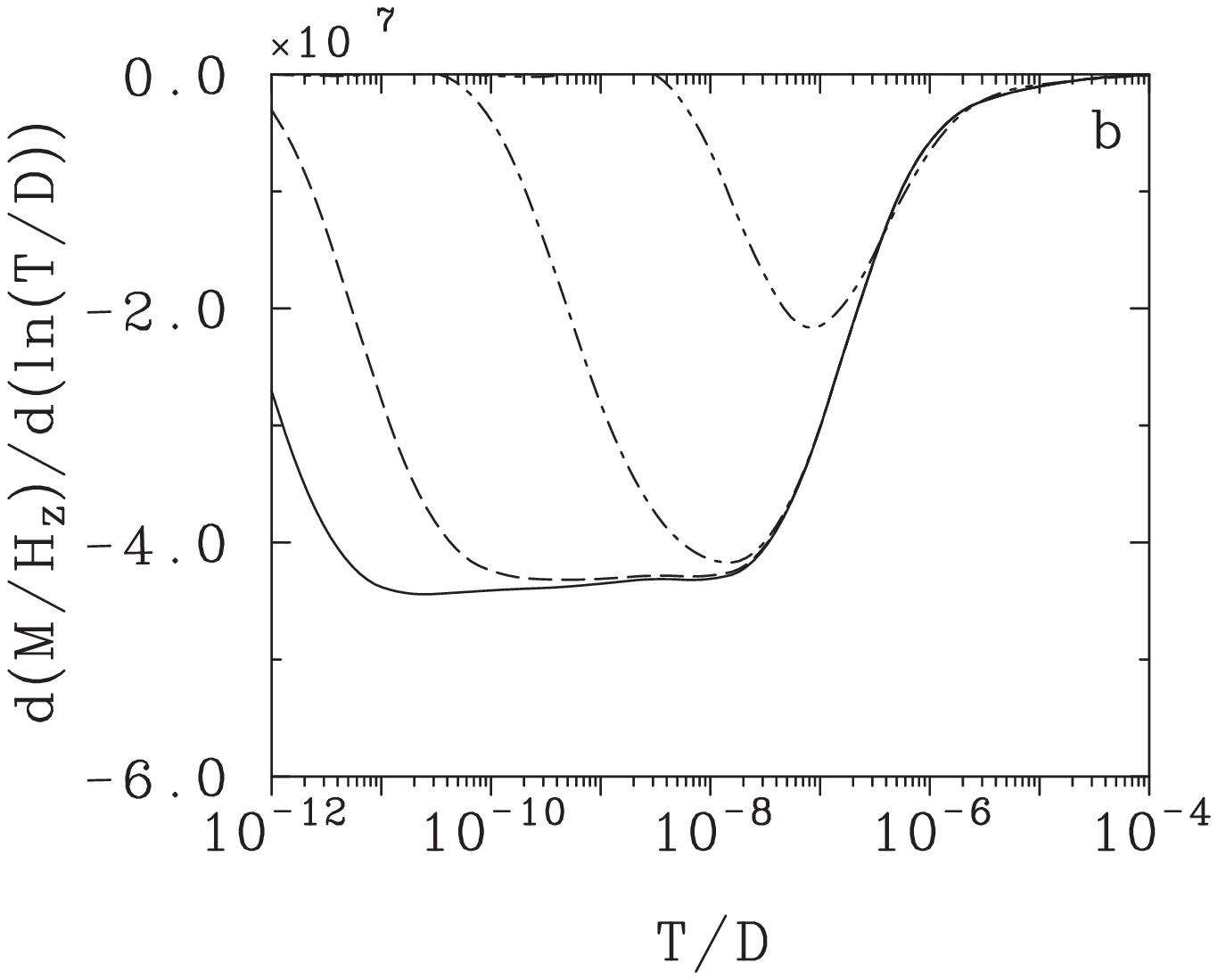}
\caption{
Temperature dependence of the magnetizations~(a)
and the differential of them
by the logarithm of temperature~(b)
under the orthorhombic CEF.
The magnetic field $H_z = 1 \times 10^{-10}$ is applied.
The split of the two singlets of f$^2$ configuration is;
$\delta = 0$~($\circ$, solid line),
$2.2 \times 10^{-9}$~($\diamond$, dashed line),
$2.1 \times 10^{-8}$~($\triangle$, dot-dashed line),
$2.1 \times 10^{-7}$~($\Box$, two-dots-dashed line), respectively.
The occupation number of the f-electron
for $\delta = 0$ is 1.91.
}
\label{fig:12}
\end{figure}

We study an effect of the lowering of the crystal symmetry as another possible
mechanism.
When the crystal symmetry changes from the tetragonal type to the orthorhombic one,
the non-Kramers doublet CEF ground state splits into two singlets.
If the split is large enough, the low energy states is expected to follow
the LFL model, and the entropy at low temperature will go to zero.
In Fig.~\ref{fig:12}~(a) we show the normalized magnetizations
for several cases of the orthorhombic CEF.
When the split, $\delta$, is zero,
the result corresponds to that for the tetragonal CEF.
Because the applied magnetic field, $H_z$, is very weak, the magnetization shows
$-\ln T$ dependence for wide temperature region;
from $T^* \sim 10^{-8}$ to $H_z / 10 \sim 10^{-11}$.
The $-\ln T$ dependence of the magnetic susceptibility is
suppressed in the low temperature region, $T < \delta/10$. 
But for the small split, the $-\ln T$ dependence is still shown
for the temperature region from $T^*$ to $\delta /10$
as shown in Fig.~\ref{fig:12}~(b).
The region is decreased with increasing $\delta$, and it disappears
for $\delta \sim T^*$.
The saturation of the magnetization occurs in the temperature range where
the energy spectrum is explained by the LFL Hamiltonian.
The residual entropy will be released at the temperature that
the magnetization saturates.
If the lowering of the crystal symmetry is not large,
the NFL behavior will be observed in a restricted temperature region.

\section{Summary}
We have investigated the NFL state of the IAM
with non-Kramers doublet ground state of the f$^2$ configuration
under the tetragonal CEF.
The low energy spectra, the temperature dependence of the magnetizations
and the magneto resistances of the model are calculated by employing the NRG method.
The spectra are explained by the combination of the NFL and the LFL parts
which are independent with each other.
One orbital of electrons with the $\Gamma_7^{(1)}$ symmetry
contributes to the spectrum
of LFL part, and the other two orbitals contribute to that of the NFL part.
The phase shift of the conduction electron for the LFL component
becomes very small at very low energy.
The NFL part of the spectra has the same form to that of $S$=1/2-TCKM
which has been derived by the CFT.

The magnetization under the weak magnetic field shows the $- \ln T$ dependence,
and it saturates below the temperature, $H_z / 10$.
The coefficient of the $-\ln T$ term has magnitude of $1/T^*$,
where $T^*$ is the temperature that the $-\ln T$ dependence begins.
The saturation is consistent with the break of the NFL state under the magnetic field.
The magnetization is well scaled by $T^*$ and $M^*$ = $M(T=T^*, H_z / T^* \ll 1)$.
The magnetic field dependence of the phase shift is calculated
from the low energy spectrum.
When the hybridization is weak,
the scattering amplitude decreases with increasing the magnetic field.
But for large hybridization, the scattering amplitude increases.
The existence of the positive magneto resistance suggests
that the resistivity decreases with decreasing temperature.
We have the parameter region for the IAM
that the $- \ln T$ term of the magnetic susceptibility has the sizable coefficient,
and the resistivity decreases with decreasing temperature.
There is the possibility that the anomalous properties of U$_x$Th$_{1-x}$Ru$_2$Si$_2$
can be explained by the NFL scenario of the TCKM type for the present
IAM.
However, Actual U ion obeys $L$-$S$ coupling scheme, so we need
to consider complicated structures of the electronic states of the multi-electron configurations for the quantitative discussions.

The lowering of the crystal symmetry breaks the NFL behavior
at the temperature about $\delta /10$,
where $\delta$ is the orthorhombic CEF splitting.
The residual entropy is released at around this temperature.
If $\delta$ is small, the NFL behavior is still
expected in the intermediate temperature range.

\section*{Acknowledgments}
We would like to thank H. Amitsuka, M. Koga, H. Shiba and Y. Kuramoto
for useful discussions and comments.
This work has been partly supported by Grant-in-Aid
Nos. 06244104, 09640451 and 09244202
from the Ministry of Education, Science and Culture of Japan.
The computation has been done at `the Computer Center, Tohoku University',
`the facilities of the Supercomputer Center, Institute for Solid State Physics,
University of Tokyo' and `the Computer Center of Institutes for Molecular Science,
Okazaki National Research Institutes'.


\begin{thebibliography}{99}
\bibitem{rf:table2} The small difference may be due to the residual small
interaction between the NFL and the LFL parts because we listed the energies at
$L=31$ step.
\end{thebibliography}
\end{document}